\documentclass[trackchanges, twocolumn]{aastex701}

\usepackage[fleqn]{amsmath}
\usepackage{multirow}
\usepackage{booktabs}
\usepackage{mathtools}
\usepackage{placeins}

\begin{document}

\title{Geant4-IcyMoons: Simulating Electron Interaction Physics in Irradiated Astrophysical Ices}

\author[orcid=0000-0002-1451-6492]{Gideon Yoffe}
\affiliation{Department of Earth and Planetary Sciences, Weizmann Institute of Science, Rehovot 76100, Israel}
\email[show]{gidi.yoffe@weizmann.ac.il}

\author[orcid=0000-0001-5830-5454]{Jacques Pienaar}
\affiliation{Department of Physics Core Facilities, Weizmann Institute of Science, Rehovot 76100, Israel}
\email{jacques.pienaar@weizmann.ac.il}

\author[orcid=0000-0003-2105-4078]{Ioanna Kyriakou}
\affiliation{Medical Physics Laboratory, Department of Medicine, University of Ioannina, Ioannina 45110, Greece}
\email{ikyriak@uoi.gr}

\author[orcid=0000-0002-9996-797X]{Dimitris Emfietzoglou}
\affiliation{Medical Physics Laboratory, Department of Medicine, University of Ioannina, Ioannina 45110, Greece}
\email{demfietz@uoi.gr}

\author[orcid=0000-0002-0619-2053]{S\'{e}bastien Incerti}
\affiliation{Université de Bordeaux, CNRS, LP2I, UMR 5797, F-33170 Gradignan, France}
\email{incerti@lp2ib.in2p3.fr}

\author[orcid=0000-0001-7068-5621]{Hoang Ngoc Tran}
\affiliation{Université de Bordeaux, CNRS, LP2I, UMR 5797, F-33170 Gradignan, France}
\email{tran@lp2ib.in2p3.fr}

\author[orcid=0000-0002-5958-9346]{Yohai Kaspi}
\affiliation{Department of Earth and Planetary Sciences, Weizmann Institute of Science, Rehovot 76100, Israel}
\email{yohai.kaspi@weizmann.ac.il}

\begin{abstract}

Energetic particles continuously process water ice across astrophysical and planetary environments---from interstellar clouds and comets to icy planetary surfaces. Interpreting the resulting observables requires a physically grounded description of the underlying interactions, both to identify radiation-driven signatures and to distinguish them from superimposed chemical, thermal, and microphysical effects. We present Geant4-IcyMoons, an extension of Geant4-DNA developed for irradiated water ice and, ultimately, for materials embedded within it. 
In this study, we model the elastic and inelastic interactions of electrons with amorphous and hexagonal ice. 
For the first time, this enables a transport-ready Monte Carlo simulation of electron irradiation in water ice, linking incident-particle environments to the evolution of icy surfaces.
We apply this framework to Jupiter's moon Europa as a representative case of electron bombardment of an icy surface. We show that, on the trailing hemisphere, the stronger low-energy electron bombardment confines much of the deposited energy to the upper $\lesssim 0.1$~cm, whereas on the leading hemisphere, the more energetic incident population drives deposition patterns to depths of tens of centimeters. This may contribute to the observed lens-like enrichment of radiolysis products centered on the equator of the trailing hemisphere. This work lays the foundation for treatments of ion irradiation and radiation chemistry in water ice and embedded materials.

\end{abstract}

\keywords{\uat{Astrophysical processes}{104} --- \uat{Radiative transfer simulations}{1967} --- \uat{Astrochemistry}{75} --- \uat{Planetary surfaces}{2113} --- 
\uat{Surface ices}{2117} --- \uat{Computational methods}{1965}}


\newpage
\section{Introduction}\label{sec:intro}

Water ice plays a central role in a wide range of astrophysical and planetary environments. Across many of these settings, it occurs predominantly in two structural forms: amorphous ice, which is favored by low-temperature deposition and irradiation, and hexagonal ice, the stable crystalline phase produced by thermal annealing or formation at higher temperatures \citep{jenniskens1994structural}. In dense molecular clouds and the interstellar medium, it coats dust grains and mediates surface chemistry that helps set molecular inventories prior to star and planet formation \citep{herbst1995chemistry,hollenbach2008water,potapov2021dust}. In protoplanetary disks, it regulates the sequestration and transport of volatiles across snowlines, thereby influencing the primordial chemical endowment of planets and small bodies \citep{lunine2006origin, ciesla2006evolution,henning2013chemistry}. Water ice is likewise a major constituent of comets \citep{snodgrass2017main}, icy satellites \citep[e.g.,][]{hansen2004amorphous}, Kuiper Belt objects \citep{brown2007collisional}, and the polar and near-surface environments of moons \citep{hayne2021micro}, planets, and dwarf planets \citep[e.g.,][]{platz2016surface,sori2019islands}, where it records irradiation, thermal evolution, volatile transport, and exchange with the surrounding space environment.

Across these settings, ice is primarily constrained through spectroscopy and thermal diagnostics \citep{boogert2015observations}. The resulting signatures reflect both the microphysics of the ice, including phase, porosity, and grain-scale structure \citep[e.g.,][]{grundy1998temperature,hansen2004amorphous,stephan2021vis}, and the presence of impurities and products of processing, including radiolysis and photolysis \citep[e.g.,][]{loeffler2006synthesis,gerakines2000carbonic,brown2013salts}. These effects are coupled: microstructure shapes energy deposition and transport, and thus influences the incorporation, retention, and spectral expression of minor species within the ice matrix \citep[e.g.,][]{he201812co2,schiltz2024characterization}.

This coupled problem is especially evident on the surfaces of icy satellites such as Jupiter's moons Europa, Ganymede, and Callisto \citep{lunine1982formation}, and Saturn's moons Enceladus and Titan \citep{lunine2017ocean}. Several such bodies are central astrobiological targets because they likely host subsurface oceans and, in some cases, oceanic contact with rocky interiors, providing liquid water and chemical disequilibria relevant to habitability \citep{pappalardo1999does,chyba2002europa, spencer2018plume}. Their surface spectra are diverse because the near-surface layer reflects both the emplacement and transport of non-ice constituents and their continuous radiation-driven modification, acting on ice whose microstructure and thermophysical state vary across terrains and hemispheres \citep[e.g.,][]{ligier2016vlt, king2022compositional, thelen2024subsurface, cartwright2025jwst, Yoffe_2026}. This diversity encodes information on surface--interior exchange and on the production, retention, and detectability of chemically informative species \citep[e.g.,][]{brown2013salts, trumbo2019h2o2, trumbo2019sodium, trumbo2023distribution, villanueva2023endogenous}.

Interpreting such observations, however, requires more than identifying absorption bands, because spectra do not map uniquely onto a single physical or chemical state of the ice. Instead, the observed signal conflates composition, microstructure, and the integrated effects of thermal evolution and radiation-driven processing \citep[e.g.,][]{trumbo2018alma,thelen2024subsurface, mergny2025blinking}. A forward model that propagates the relevant radiation environment through heterogeneous ice and predicts the resulting energy-deposition field would therefore provide a more direct link between observables and the processes that reshape the measurable near-surface layer \citep[e.g.,][]{fama2010radiation,mitchell2017porosity,loeffler2020possible, mergny2024lunaicy}. Such a framework is central for constraining volatile budgets and isotopic signatures \citep[e.g.,][]{gerakines2000carbonic,he201812co2}, and for assessing the detectability and survivability of organic tracers under realistic irradiation histories \citep{nordheim2018preservation, pavlov2024radiolytic, yoffe2025fluorescent}.

Implementing such a mapping requires addressing a high-dimensional parameter space spanning particle spectra and fluxes \citep[e.g.,][]{fama2010radiation,loeffler2020possible}, impurity chemistry and mixing state \citep[e.g.,][]{brown2013salts,trumbo2019sodium}, and the temperature- and microstructure-dependence of transport and kinetics \citep[e.g.,][]{grundy1998temperature,mitchell2017porosity,stephan2021vis}. It must also capture the time-dependent coupling among radiolysis, photolysis, and thermal reprocessing \citep{mergny2025blinking}, and ultimately depends on laboratory programs spanning mission-relevant ranges of temperature, composition, morphology, and radiation history, together with radiative-transfer and radiation-chemistry models that connect laboratory constraints to observable spectra \citep[e.g.,][]{strazzulla1992ion,loeffler2020possible,he2022refractive}.

Monte Carlo particle transport provides a natural framework for linking microscopic interaction physics to macroscopic observables. By following the stochastic sequence of elastic and inelastic interactions experienced by individual particles, one may derive energy-deposition fields, secondary-particle production, and spatial dose distributions across a wide range of materials and incident energies. The Geant4 toolkit, developed at CERN, is the standard platform for detector simulation and radiation transport in accelerator-based experiments \citep{agostinelli2003geant4, allison2006geant4}, with extensive validation and broad adoption \citep{allison2016recent}. In parallel, track-structure Monte Carlo methods have been developed in radiobiology and medical physics to model low-energy electrons in condensed media, where nanometer-scale energy deposition governs chemical damage. Geant4-DNA \citep{incerti2010geant4, incerti2010comparison, bernal2015track, incerti2018geant4,kyriakou2021review, tran2024review} provides the most comprehensive open framework bridging these regimes, combining a mature transport infrastructure with condensed-matter interaction models. However, it is primarily optimized for liquid water at room temperature and is therefore not directly suited to cold, condensed-phase water ice \citep{incerti2018geant4}.

Here, we introduce Geant4-IcyMoons, an extension of Geant4-DNA designed to simulate coupled physical and later chemical interactions between radiation and water ice, including embedded materials. We construct a transport-ready description of electron interactions in solid ice, spanning elastic scattering and inelastic channels that partition energy into vibrational, electronic, and ionization losses. This requires differential cross sections defined over the kinematic phase space sampled by particle transport, parameterized by incident energy, energy transfer, and scattering kinematics, namely the angle (or, equivalently, the momentum transfer). In this work, we denote these quantities by $T$, $E$, $\theta$ (or $\Omega$ for solid angle), and $q$, respectively.

In condensed ice, no single closed-form model maintains comparable accuracy across all interaction channels, because the relevant response changes qualitatively between regimes. Low-energy losses are structured by discrete, phase-dependent modes \citep{sanche1995interactions}, whereas higher-energy losses are governed by collective, screened electronic response \citep{emfietzoglou2003inelastic,emfietzoglou2013inelastic}. Experimental constraints are likewise heterogeneous, since different channels are accessed through different observables and do not provide uniform coverage in energy or phase \citep[e.g.,][]{michaud2003cross,shinotsuka2017calculations,signorell2020electron}. We therefore adopt a hybrid construction: experimentally derived cross sections are used where available, and otherwise we apply physically constrained extensions that recover the correct asymptotic behavior and yield self-consistent transport kinematics. In the present implementation, the distinction between amorphous and hexagonal ice arises primarily through the electronic excitation and ionization channels, whose dielectric responses are phase-resolved. The remaining channels are treated, to first order, using amorphous-ice constraints, since comparable phase-resolved data are not yet available. The resulting differences between amorphous and hexagonal ice should therefore be interpreted mainly as consequences of their distinct electronic inelastic response, while possible phase-dependent corrections to elastic, vibrational, and attachment processes are deferred to future work. The combined model is valid over the energy range 1--10$^{7}$ eV.

This paper is organized around the construction, validation, and application of Geant4-IcyMoons. We first develop the physical interaction model for electrons in condensed water ice, beginning with elastic scattering across the low- and high-energy regimes (Section~\ref{sec:elastic}), then introducing vibrational excitation in amorphous ice (Section~\ref{sec:vib}) and low-energy attachment processes (Section~\ref{sec:deattach}). We next present the dielectric-response framework used to model electronic excitation and ionization (Section~\ref{sec:excitation_ionization}), thereby completing the transport description of electron interactions in ice. We then assess the model through benchmarking and validation tests (Section~\ref{sec:benchmarking}) and specify how transport is handled outside the formal range of the present implementation (Section~\ref{sec:outofrange}). With that framework in place, we apply Geant4-IcyMoons to electron bombardment of Europa as a representative use case (Section~\ref{sec:Europa}), and conclude by summarizing the main results and their implications (Section~\ref{conclusions}).

\section{Elastic scattering}\label{sec:elastic}

Elastic scattering describes the deflection of an incident electron by the electrostatic potential of atoms or molecules. Although a small recoil energy can, in principle, be transferred to the target, this contribution is negligible in condensed media because the target is much more massive than the electron. Elastic scattering is therefore treated, to good approximation, as changing only the electron direction of motion \citep[e.g.,][]{ptasinska2022electron}.
It is one of the dominant processes governing electron transport in condensed media, particularly at energies below a few hundred~eV where multiple scattering strongly influences the spatial evolution of particle tracks \citep{nikjoo2016radiation}. 
In this regime, elastic events control the angular distribution, shaping the local energy deposition profile that drives radiolytic chemistry and secondary-electron cascades.  
At higher energies, elastic scattering becomes increasingly forward-peaked and primarily contributes to gradual angular diffusion, whereas at low energies it determines the pattern of electronic energy transfer.  

In the standard Geant4 electromagnetic physics, electron elastic scattering is treated through a screened Rutherford model, which represents a two-body interaction between a free electron and an isolated atomic nucleus shielded by its surrounding electron cloud \citep{mendenhall2005algorithm}. This formulation is based on the Wentzel potential, describing scattering in a screened Coulomb field that accounts for the attenuation of the nuclear charge by atomic electrons.
This model is computationally efficient and has been widely adopted in Monte Carlo transport codes. It performs well for high-energy electrons and dilute gases, where scattering centers act independently, but becomes inaccurate in condensed phases where coherence, exchange, and collective screening effects are significant \citep{nikjoo2016radiation}.

In Geant4-DNA, elastic scattering of electrons in liquid water is implemented through several models \citep{incerti2010comparison, bernal2015track, incerti2018geant4, tran2024review}, of which the most recent employs the \textit{Elastic Scattering of Electrons and Positrons by Atoms} (ELSEPA) code to compute differential elastic cross-sections for electrons interacting with water in the condensed phase \citep{shin2018development}.  
ELSEPA is a relativistic partial-wave solver that integrates the Dirac equation for a given atomic or molecular potential, accounting for exchange and correlation effects \citep{salvat2005elsepa}.  
This implementation extends the valid energy range down to approximately 10~eV and provides an accurate treatment of low-energy scattering via numerical phase-shift analysis, yielding the most physically consistent description of electron elastic scattering in liquid water currently available. 

Here, we adopt the empirical elastic cross-sections for amorphous ice measured over the energy range 1–100 eV, obtained from a two-stream multiple-scattering analysis of electron backscattering spectra that explicitly incorporates condensed-phase interference effects \citep{michaud2003cross}. These cross-sections predominantly reflect isotropic scattering, which suppresses coherent forward scattering in amorphous ice, and provide a physically realistic description of condensed water that serves as a foundation for modeling elastic interactions in both amorphous and crystalline ice.
Previous work has shown that elastic cross-sections derived for the gas phase lead to unrealistically short mean free paths below a few tens of electron volts because the Born approximation and isolated-atom potentials overestimate forward scattering and neglect collective screening \citep{nikjoo2016radiation}. Within the same review, the amorphous-ice cross-sections reported in \citet{michaud2003cross} were considered too small, underestimating large-angle scattering and yielding mean free paths likely longer than those of liquid water; scaling factors of 2 to 6 were therefore advised to correct this discrepancy. Subsequent experimental studies have revisited this interpretation. Subsequent measurements of electron attenuation lengths in liquid water and amorphous ice have challenged that interpretation and argue against applying any such scaling factor \citep{signorell2020electron}. 

Building on this evidence, we employ the condensed-phase cross-sections without additional scaling and smoothly extend them to higher energies, ensuring consistency with the ELSEPA single-atom treatment. The aim is to provide a continuous connection between the low-energy condensed-phase regime and the high-energy independent-atom regime.
To determine the transition between condensed-phase and gas-like behavior, we adopt a physically motivated criterion based on the relation between the electron de~Broglie wavelength and the average distance between neighboring oxygen atoms in the hydrogen-bonded lattice, known as the intermolecular O--O separation.  
The O--O separation reported in the measurements for amorphous water ice is $d \approx 2.76$~\AA \citep{michaud2003cross}.  
When the electron wavelength becomes much smaller than this intermolecular spacing, its wave function can no longer experience coherent scattering from neighboring molecules, and individual scattering centers act approximately independently.  
We therefore define the transition kinetic energy $T_t$ by the condition $\lambda/d < 0.2$, where $\lambda = h / \sqrt{2 m_e T}$ is the electron de~Broglie wavelength and $T$ is its kinetic energy.  
For $d = 2.76$~\AA, this yields $\lambda = 0.552$~\AA\ at $\lambda/d = 0.2$, corresponding to $T_t \approx 494$~eV.  
At this energy, elastic scattering in the condensed phase asymptotically approaches the single-atom description. 

\begin{figure}[h!]
\centering
\rotatebox[origin=c]{0}{\includegraphics[scale = 0.32]{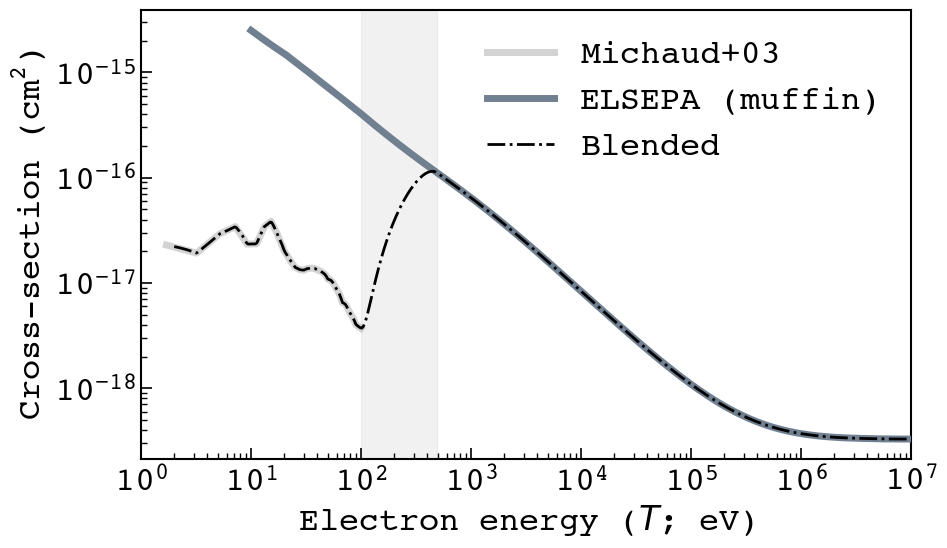}}
  \caption{Total cross-sections for elastic scattering. The light gray curve shows the 1--100~eV measurements of \citet{michaud2003cross}. The slate gray curve shows the cross-sections of the ELSEPA model for liquid water. The dashed black curve shows the blended elastic model adopted in Geant4-IcyMoons.}
     \label{fig:elastic_cs}
\end{figure}

To ensure a continuous transition between the empirical amorphous-ice data and the high-energy single-atom regime, we employ a smooth interpolation kernel defined as a cubic smoothstep function:

\begin{equation}
S(T)=3s^{2}-2s^{3},\qquad
s=\left.\frac{T-100~\mathrm{eV}}{T_t-100~\mathrm{eV}}\right|_{0}^{1},
\end{equation}

where \(s\) is the normalized interpolation variable, clipped to the interval \([0,1]\). Thus, \(s=0\) for \(T\le 100~\mathrm{eV}\), \(s=1\) for \(T\ge T_t\), and \(0<s<1\) only within the transition region \(100~\mathrm{eV}<T<T_t\). The blended elastic cross-section is then given by  

\begin{equation}
\begin{aligned}
\sigma_{\mathrm{blend}}(T)
&=\bigl[1-S(T)\bigr]\,\sigma_{\mathrm{ice}}(T)\\
&\quad+S(T)\,\sigma_{\mathrm{ELSEPA}}(T)\,.
\end{aligned}
\end{equation}

where $\sigma_{\mathrm{ice}}$ represents the empirical amorphous-ice cross-section and $\sigma_{\mathrm{ELSEPA}}$ the ELSEPA cross-section.
This formulation preserves the condensed-phase behavior below 100~eV, smoothly merges into the classical single-atom regime near 494~eV, and ensures continuity in both magnitude and slope across the transition (see Figure~\ref{fig:elastic_cs}).  
The corresponding angular distributions are merged by sampling deflections isotropically in the empirical branch up to 200~eV, and above that threshold from tabulated differential cross sections of the ELSEPA model, which combines the low-energy condensed-phase constraints with a forward-peaked single-atom description.

\section{Vibrational excitation}\label{sec:vib}

\begin{figure}[t!]
\centering
\rotatebox[origin=c]{0}{\includegraphics[scale = 0.32]{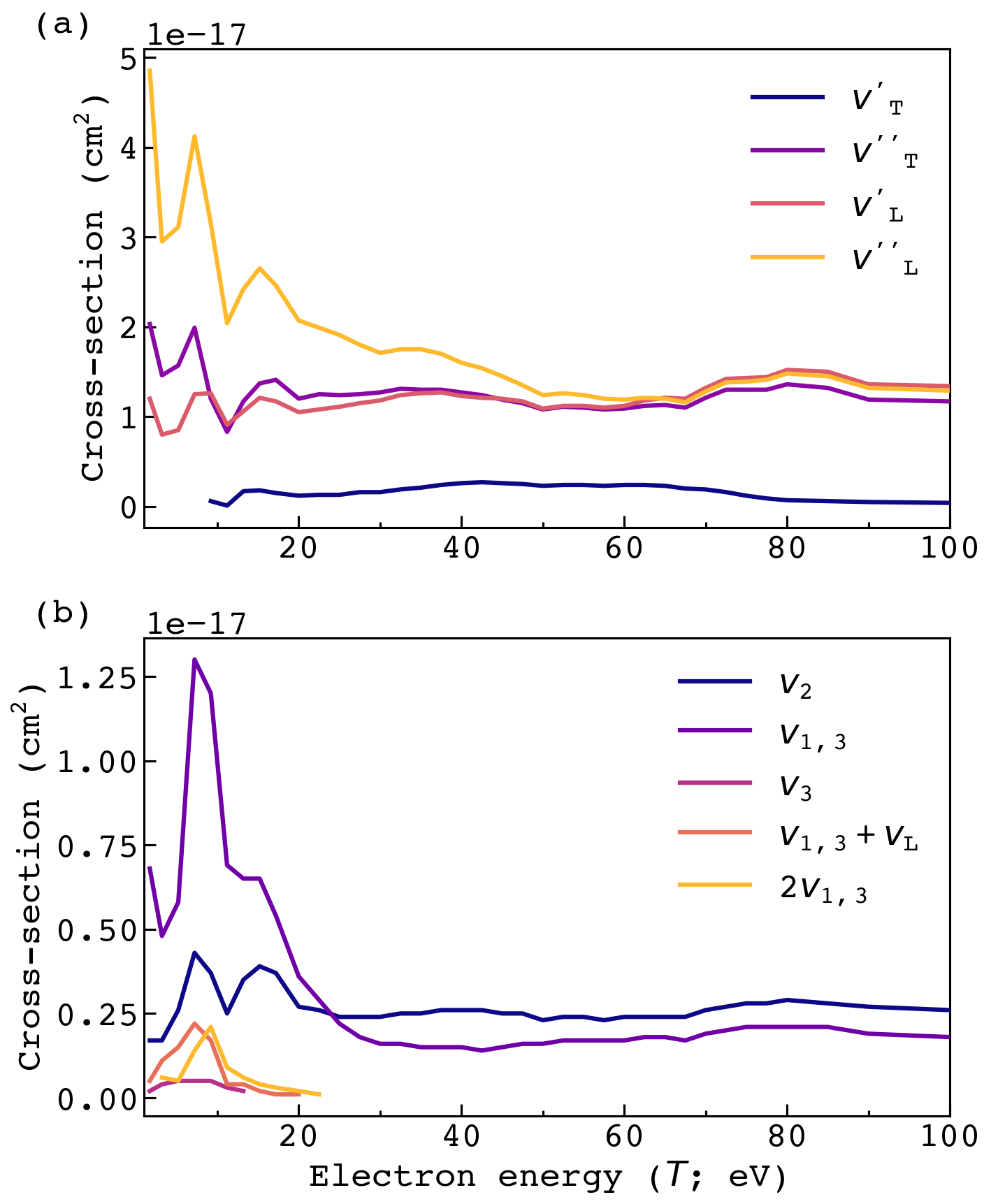}}
  \caption{Total cross-sections of vibrational excitation channels, as reported in \citet{michaud2003cross} for (\textbf{a}) intermolecular- and (\textbf{b}) intramolecular- modes.}
     \label{fig:vib_cs}
\end{figure}

Vibrational excitation represents a dominant inelastic channel for sub-20~eV electrons in condensed water, mediating the transfer of electronic energy into the molecular lattice.  
In amorphous ice, these processes encompass both \emph{intermolecular} and \emph{intramolecular} modes, each corresponding to distinct collective or molecular degrees of freedom.
The intermolecular modes include four low-energy excitations associated with the hydrogen-bond network: two translational (phonon) modes, $v'_{\mathrm{T}}$ and $v''_{\mathrm{T}}$ at approximately 10 and 24~meV, and two librational modes, $v'_{\mathrm{L}}$ and $v''_{\mathrm{L}}$ centered near 61 and 92~meV.  
These excitations arise from hindered translations and rotations of water molecules within the disordered lattice, representing the quantized vibrations of the condensed network.
The intramolecular modes comprise five higher-energy vibrations intrinsic to the $\mathrm{H_2O}$ molecule: the bending mode $v_2$ near 0.20~eV; the symmetric and asymmetric stretching modes $v_1$ and $v_3$ at 0.42--0.47~eV; the stretch–libration combination band $v_{1,3}+v_{\mathrm{L}}$ around 0.51~eV; and the overtone $2v_{1,3}$ near 0.83~eV.  
These nine channels were experimentally resolved by \citet{michaud2003cross} through high-resolution electron energy-loss spectra of amorphous ice at 14 K, modeled using a two-stream multiple-scattering formalism that distinguishes between isotropic and forward components of the scattering probability.  
The resulting integral cross-sections quantify how incident electrons deposit quantized energy into phonons and molecular vibrations, linking microscopic excitation probabilities to macroscopic energy dissipation in irradiated ice (see Figure~\ref{fig:vib_cs}).  

In their implementations within Geant4-IcyMoons, these vibrational channels are restricted to incident electron energies below 100~eV, above which the scattering probability for phonon and molecular vibrational modes becomes negligible compared to electronic excitation and ionization processes \citep{michaud2003cross}.

\subsection{Energy-loss distributions}\label{sec:vib_energyLoss_dist}

For each inelastic vibrational event, an incident electron of kinetic energy $T$ transfers an energy $E$ to the surrounding medium.
In amorphous ice, the quantized vibrational losses are broadened by variations in local bonding and structural disorder.
Following \citet{michaud2003cross}, each vibrational channel is therefore represented by a Gaussian distribution in the energy-transfer variable $E$, centered at a mean loss $E^{(\rm v)}_i$ and with full width at half maximum (FWHM) $\Delta_i$.

The normal energy-loss distribution for the $i^{\rm{th}}$ mode is
\begin{equation}
\mathcal{N}_i(E; E^{(\rm v)}_i, b_i) = \frac{1}{\sqrt{\pi}\,b_i}
\exp\!\left[-\frac{(E - E_i^{(\rm v)})^2}{b_i^2}\right],
\end{equation}
where $b_i = \Delta_i / [2\sqrt{\ln 2}]$.
The resulting vibrational differential cross-section in transferred energy is then written as a weighted sum of channels,
\begin{equation}
\frac{d\sigma}{dE}(T; E) = \sum_i \sigma_i(T)\, \mathcal{N}_i(E; E^{(\rm v)}_i, b_i),
\end{equation}
where $\sigma_i(T)$ is the integral cross-section for $i^{\rm{th}}$ mode at incident energy $T$.
This parameterization preserves the quantized structure of the vibrational spectrum while reproducing the experimentally observed broadening, enabling direct sampling of energy losses in Monte Carlo transport.
The measured values of $E^{(\rm v)}_i$ and $\Delta_i$ for each channel are listed in Appendix~\ref{app:vibparam}.

In the original Geant4-DNA implementation, vibrational channels were treated as discrete inelastic processes with fixed energy transfers $E^{(\rm v)}_i$.
To reproduce the experimental broadening reported by \citet{michaud2003cross}, the Geant4-IcyMoons implementation samples the transferred energy of each vibrational event from the normal distribution $\mathcal{N}_i(E; E^{(\rm v)}_i, b_i)$ corresponding to that event.

\subsection{Angular deflection and anisotropy}\label{sec:vib_angles}

In addition to energy loss, each inelastic event involves a change in the electron’s direction of motion determined by the angular distribution of scattering.  
In \citet{michaud2003cross}, this behavior was characterized using an anisotropy coefficient $\gamma_i(E_0)$, which defines the relative strength of forward and backward scattering for each excitation mode at a given incident energy.  
Within their two-stream formalism, the total scattering probability per unit path length $Q_t$ is partitioned into an isotropic component $Q_r = (1-\gamma_i)Q_t/2$ and a forward-directed component $Q_f = (1+\gamma_i)Q_t/2$.  
Small $\gamma_i$ values correspond to nearly isotropic scattering, while $\gamma_i \to 1$ indicates strongly forward-peaked behavior.
To implement this empirical anisotropy in Geant4-IcyMoons, we map it onto a continuous angular kernel using the Henyey--Greenstein (HG) phase function, which is widely used in radiative and electron-transport modeling \citep[e.g.,][]{afanas2019application}. For the $i^{\mathrm{th}}$ vibrational mode, the angular probability density is
\begin{equation}
f_i(\theta; g_i) = \frac{1-g_i^2}{4\pi\,[1+g_i^2-2g_i\cos\theta]^{3/2}},
\end{equation}
where $g_i = \langle \cos\theta \rangle$ is the asymmetry parameter.  
The HG kernel reproduces isotropic scattering for $g_i = 0$ and becomes increasingly forward-peaked as $g_i \rightarrow 1$.  
To ensure consistency with the experimental anisotropy, $g_i$ is numerically determined from the relation
\begin{equation}
\int_0^{\pi/2} f_i(\theta; g_i)\, 2\pi \sin\theta\, d\theta = \frac{1+\gamma_i(E_0)}{2},
\end{equation}
so that the forward-to-backward scattering ratio exactly matches the values reported by \citet{michaud2003cross}.

In Geant4-IcyMoons, each $i^{\rm{th}}$ vibrational excitation channel is assigned its own energy-dependent $g_i(E_0)$, tabulated from the corresponding $\gamma_i$ values.  
During simulation, the scattering angle $\theta$ is sampled from the cumulative distribution of the HG function using inverse-transform sampling, ensuring that the generated deflection angles reproduce the empirical anisotropy for each vibrational mode.  
This implementation replaces the original two-stream (hemispheric) approximation with a smooth, continuous angular kernel, providing a more realistic description of angular diffusion in amorphous ice. A derivation of the numerical procedure is provided in Appendix~\ref{app:HG_numerical}.

\section{Dissociative attachment and detachment}\label{sec:deattach}

At very low electron energies ($\lesssim$6~eV), slow electrons can be captured by water molecules, forming transient negative-ion states. In this regime, dissociative electron attachment becomes an efficient interaction channel \citep{sanche1995interactions,bass2003dissociative}. These predominantly relax via bond cleavage, converting the electron’s kinetic energy into localized excitation and chemical modification rather than returning it to the transport cascade.
In the laboratory, measurements of water ice showed that dissociative attachment could not be cleanly separated from other sub-excitation inelastic processes. Instead of treating attachment as a resolved reaction pathway, its contribution was incorporated into the total low-energy inelastic response of amorphous ice, leading to the experimentally observed enhanced attenuation of electrons below a few electron volts \citep{michaud2003cross}.

\begin{figure}[h!]
\centering
\rotatebox[origin=c]{0}{\includegraphics[scale = 0.32]{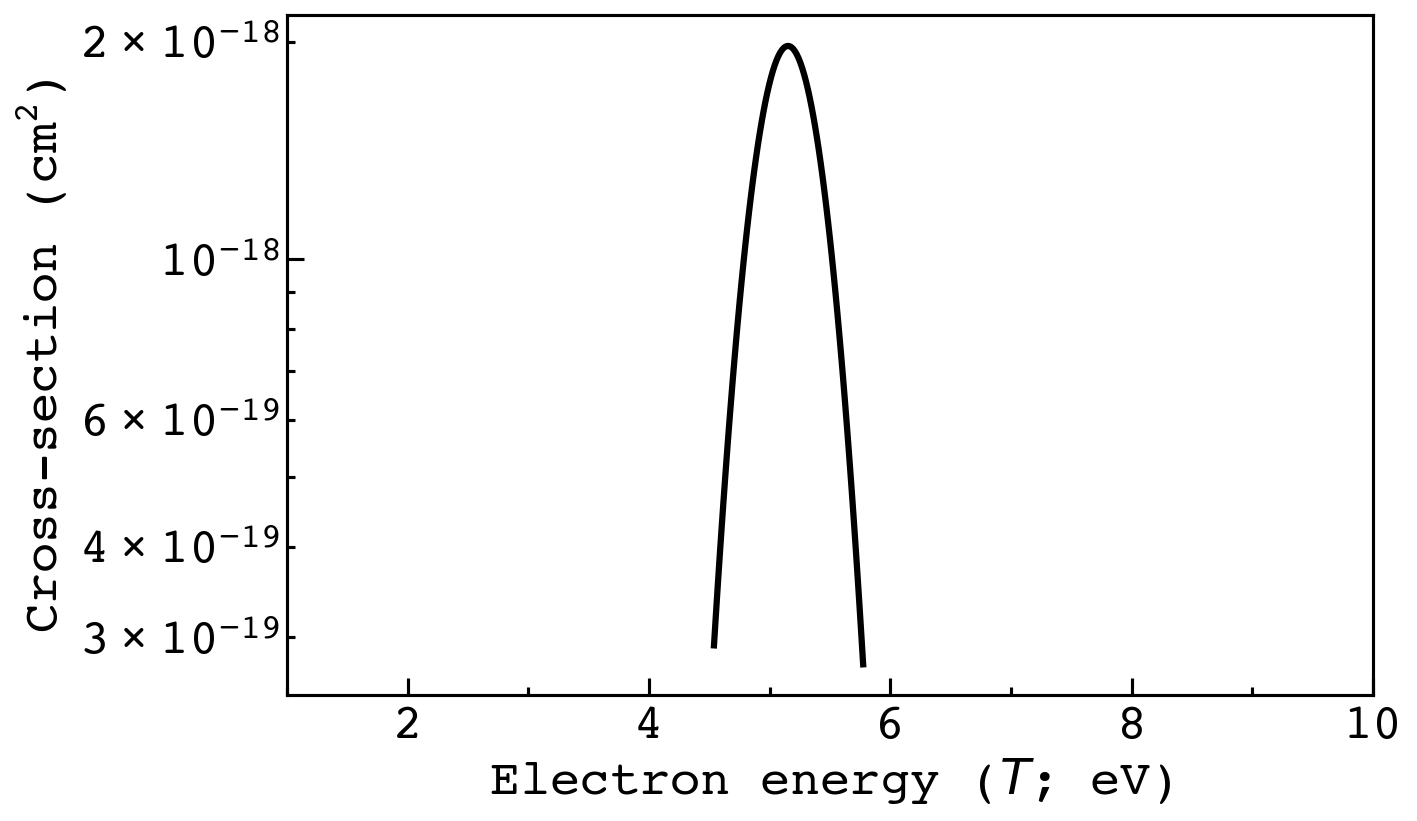}}
  \caption{Experimentally-derived total dissociative electron attachment cross sections for amorphous water ice \citep{michaud2003cross}.}
     \label{fig:attach}
\end{figure}

Geant4-DNA, and by extension Geant4-IcyMoons, adopts this experimental perspective. Dissociative attachment is implemented as an effective capture channel defined directly in terms of the incident electron kinetic energy, with kinetic-energy-dependent cross-sections derived from \citet{michaud2003cross}. The process is confined to a narrow energy interval, approximately 4.5–5.8~eV, with a single maximum near 5.1~eV (see Figure~\ref{fig:attach}). When an attachment event occurs, the electron is absorbed, and its remaining kinetic energy is deposited locally, terminating the transport track. If chemical processes are enabled, a dissociative-attachment product is recorded. Electron detachment is not treated explicitly.
This approach captures the experimentally constrained role of dissociative attachment as a sink for low-energy electrons in amorphous ice, without introducing assumptions about transient anion lifetimes or decay pathways that are not resolved by available measurements.

\section{Electronic excitation and ionization}\label{sec:excitation_ionization}

Beyond the vibrational domain, electrons interacting with water ice can transfer greater amounts of energy to the medium's electronic subsystem, opening channels for electronic excitation and ionization.  
These processes occur once the transferred energy $E$ exceeds the upper bound of the vibrational spectrum \citep[$\mathcal{O}(10)$~eV;][]{emfietzoglou2007consistent}, engaging the valence and inner electronic states of the water molecule.  

Each excitation or ionization channel represents a distinct mode of energy loss, defined by its threshold and spectral shape.  
In practice, the electronic inelastic response of water and ice is represented by five major excitation channels (indexed hereafter by $j=1$--5) and four ionization channels (indexed hereafter by $i=1$--4). The excitation channels are modeled as five discrete bands in the valence-loss region, typically spanning $\sim$8--15~eV, whereas the ionization channels correspond to the four principal valence subshells of water \citep{emfietzoglou2002inelastic, emfietzoglou2005complete}. 
These include the $A_1 \!\leftrightarrow\! B_1$ excitations, which label low-energy excitation features in the optical model, as well as two Rydberg-like features corresponding to excitations into spatially extended, weakly bound electronic states (labeled A+B and C+D in the optical data models).  
The final component is a broad diffuse band attributed to overlapping higher-lying resonances and delocalized electronic motion within the hydrogen-bonded network \citep{emfietzoglou2007consistent}.  

The interaction of an electron with a condensed medium is most naturally described in the energy--momentum plane, using the variables $E$ and $q$ (see Section~\ref{sec:intro}). The energy transfer $E$ determines the type of excitation produced (vibrational, electronic, or ionizing), whereas the momentum transfer $q$ sets its spatial scale. 
Here, we describe the excitation and ionization response of the medium within the \emph{dielectric} formalism, under the assumption that it is homogeneous and isotropic. In this framework, the inelastic response is represented by the complex dielectric function, $\varepsilon(E,q)$, defined over the energy--momentum plane. The corresponding energy-loss function (ELF),
\begin{equation}
\mathrm{ELF}(E,q)=\Im\!\left[-\frac{1}{\varepsilon(E,q)}\right],
\end{equation}
gives the probability density for a single inelastic event with energy transfer $E$ and momentum transfer $q$. In what follows, we distinguish between the small-$q$ regime, where the response is controlled by long-wavelength collective screening, and the finite-$q$ regime, where the interaction becomes progressively more local, and the excitation spectrum disperses and broadens.

\subsection{Dielectric response at the optical limit}\label{sec:excitation_ionization_optical}

When the transferred momentum is small ($q\to 0$), the perturbation induced by the electron has a wavelength $\lambda = 2\pi/q$ much larger than the intermolecular spacing ($\sim$2.8~\AA\ in amorphous ice; see Section~\ref{sec:elastic}). In this long-wavelength regime, the medium may be treated approximately as a homogeneous polarizable continuum. The electron then couples primarily to the collective electronic response rather than to spatially localized scattering centers. This regime is known as the \emph{optical limit}, and corresponds to the same response probed by optical absorption and reflection experiments.
In the optical limit, the dielectric response reduces to $\varepsilon(E,0)$, and the corresponding optical ELF is
\begin{equation} \label{eq:elf}
\mathrm{ELF}(E,q{=}0)=\Im\!\left[-\frac{1}{\varepsilon(E,0)}\right].
\end{equation}

Following \citet{emfietzoglou2007consistent}, we describe $\varepsilon(E,0)$ for amorphous and hexagonal ice using an analytic representation based on a superposition of Drude-type oscillators fitted to the measured optical data. The dielectric function is written as
\begin{equation} \label{eq:varepsilon}
\varepsilon(E,0) = \varepsilon_1(E,0) + i\,\varepsilon_2(E,0),
\end{equation}
where $\varepsilon_1$ and $\varepsilon_2$ are the real and imaginary components, respectively. The imaginary part, $\varepsilon_2$, characterizes the absorptive response of the medium: it describes the out-of-phase component of the polarization and therefore the portion of the interaction in which energy is irreversibly transferred from the incident electron to the material. The real part, $\varepsilon_1$, describes the dispersive response, namely the component of the polarization that stores and releases energy reversibly and thereby sets the propagation and screening of the perturbation.
In the Drude formalism, each excitation or ionization channel contributes to the imaginary part of the dielectric function according to

\begin{equation} \label{eq:epsilon2}
\begin{aligned}
\varepsilon_2(E; q{=}0)
&=E_p^2\Biggl[
\sum_{i=1}^{4} D_2\!\bigl(E; f_i, E_{0,i}, \Gamma_i\bigr)\\
&\qquad\qquad+\sum_{j=1}^{5} D'_2\!\bigl(E; f_j, E_{0,j}, \Gamma_j\bigr)
\Biggr],
\end{aligned}
\end{equation}

where $E_p$ is the nominal free–electron plasma energy of the medium ($E_p = 20.82$ and $20.59~\mathrm{eV}$ for amorphous and hexagonal ice, respectively; \citet{emfietzoglou2007consistent}), $f$ is the oscillator strength, $E_{0,i}$ and $E_{0,j}$ are the central ionization and excitation energies, respectively, and $\Gamma$ is the damping width representing the finite lifetime of each transition. 
The real part of the dielectric function is obtained as

\begin{equation}
\begin{aligned}
\varepsilon_1(E; q{=}0)
&=1+E_p^2\Biggl[
\sum_{i=1}^{4} D_1\!\bigl(E; f_i, E_{0,i}, \Gamma_i\bigr)\\
&\qquad\qquad+\sum_{j=1}^{5} D'_1\!\bigl(E; f_j, E_{0,j}, \Gamma_j\bigr)
\Biggr].
\end{aligned}
\end{equation}

\begin{figure*}[t!]
\centering
\rotatebox[origin=c]{0}{\includegraphics[scale = 0.45]{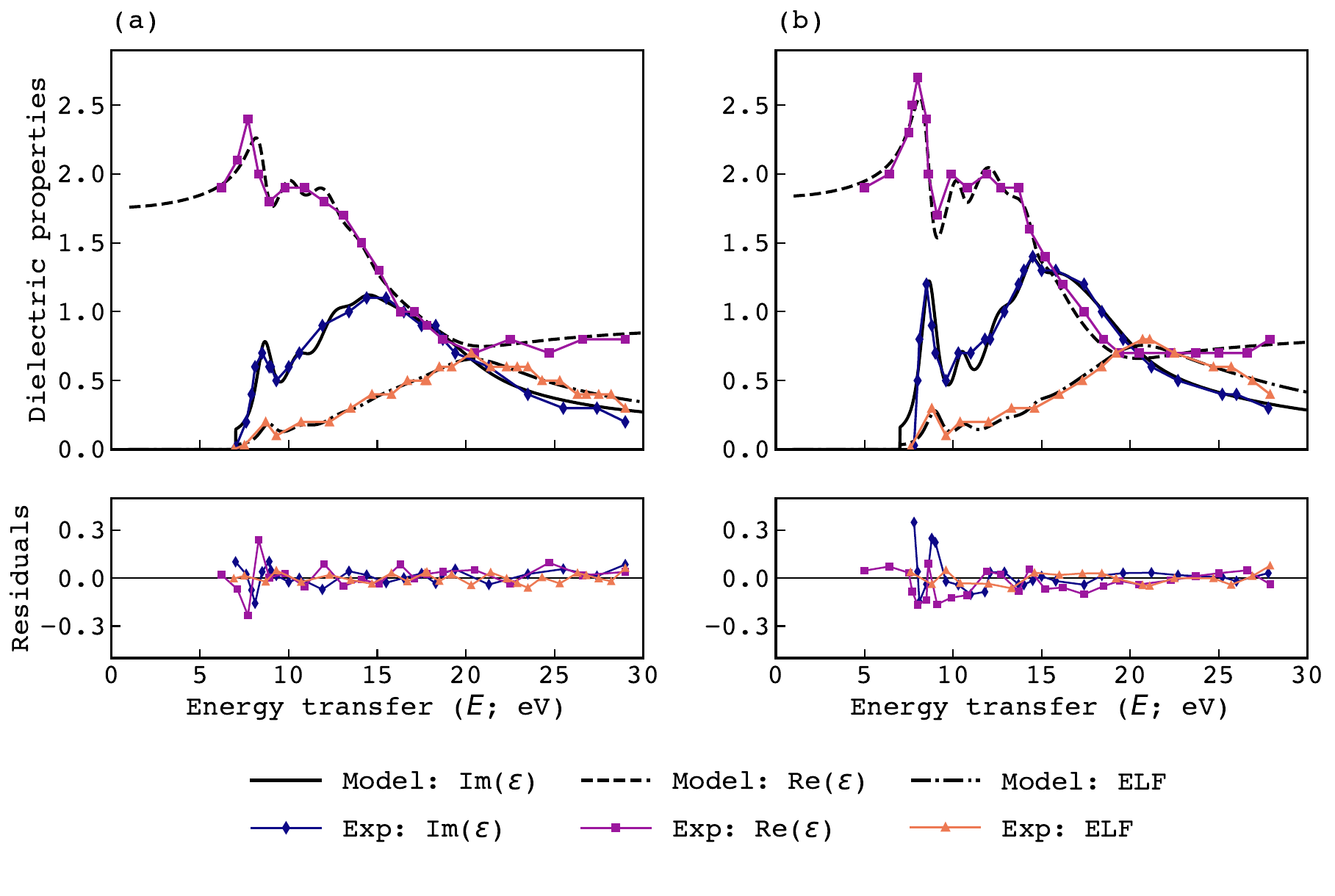}}
  \caption{Dielectric properties of amorphous and hexagonal ice in the optical limit. \textbf{Upper panels}: Best-fit Drude representations of the dielectric properties of (\textbf{a}) amorphous and (\textbf{b}) hexagonal ice at optical frequencies, based on \citet{emfietzoglou2007consistent}, compared with experimental results \citep{daniels1971bestimmung, kobayashi1983optical}. \textbf{Bottom panels}: Residuals between the best-fit Drude model and optical experimental data.}
     \label{fig:dielectric_comparison}
\end{figure*}

The ELF entering the model is constrained in the optical limit, where it is directly related to the dielectric response measured by optical spectroscopy \citep{daniels1971bestimmung, kobayashi1983optical}. This optical-limit response provides the empirical anchor for extending the dielectric model across the full energy--momentum plane, as discussed below. Figure~\ref{fig:dielectric_comparison} shows a comparison between Drude models for amorphous and hexagonal ice, based on the fitted parameters of \citet{emfietzoglou2007consistent}, with the experimental data. Lastly, a correction to the excitation--ionization partitioning was required to ensure that truncated ionization strength is reassigned only to physically permitted excitation channels \citep{kyriakou2015improvements}. The revised scheme uses $B_1$ only to delimit redistribution, and $B_{\min}$ only to gate the excitation sector, thereby preventing unphysical mixing between the two regions. The properties of the individual Drude kernels for the excitation and ionization channels, together with a description of this correction and its implementation, are provided in Appendix~\ref{app:excit_ion}.

\subsection{Finite momentum response}\label{sec:excitation_ionization_finite-q}

For finite momentum transfer ($q>0$), the electron couples to the medium on shorter length scales, because the perturbation wavelength $\lambda\sim 2\pi/q$ decreases with increasing $q$, and the energy-loss spectrum therefore broadens and disperses as the response shifts from collective screening toward more local, single-particle excitations.
Starting from the optical fit $\varepsilon(\omega,0)$, we extend the dielectric function to $q>0$ using the extended Drude formalism combined with the \emph{Emfietzoglou--Cucinotta--Nikjoo} (ECN) dispersion scheme \citep{emfietzoglou2005complete}. The ECN parameterization reproduces the observed energy–momentum dependence of the ELF of condensed water \citep{emfietzoglou2007consistent, emfietzoglou2017monte}.  
It introduces phenomenological dispersion laws for the oscillator strengths, critical energies, and damping coefficients, constrained by inelastic X-ray scattering measurements of liquid water \citep{hayashi2000complete}. For water ice, we adopt the same dispersion scheme, while fitting the optical-limit dielectric response separately for amorphous and hexagonal ice (see Figure~\ref{fig:dielectric_comparison}).
We write $q$ in atomic units $a_0^{-1}$, where $a_0$ is the Bohr radius, and use $\mathrm{Ry}=13.6057$~eV, so that $\mathrm{Ry}\,q^2$ carries the free-electron kinematic scaling along the Bethe ridge, which is the locus of maximum scattering intensity where $E\!\approx\!\mathrm{Ry}\,q^2$, reflecting the crossover toward single-electron response \citep{bethe1930theorie}.

Excitation channels are taken to be nondispersive in energy, $E_j(q)=E_j$, but their oscillator strengths and widths vary with momentum transfer. Ionization channels, by contrast, are allowed both to shift in energy and to broaden with $q$. With $(f_j,E_j,\gamma_j)$ denoting the optical-limit parameters of the five excitation bands, and $(f_i,E_i,\gamma_i)$ those of the four valence-ionization bands, the ECN dispersion scheme is
\begin{subequations}\label{eq:ecn_dispersion}
\begin{align}
f_{j}(q)
&= f_{j}\!\left[\exp\!\left(-a_{j}q^{2}\right)
+ b_{j}q^{2}\exp\!\left(-c_{j}q^{2}\right)\right], \\[4pt]
f_{i}(q)
&= f_{i}\,
\frac{1-\sum_{j=1}^{5} f_{j}(q)}{1-\sum_{j=1}^{5} f_{j}(0)}, \\[4pt]
E_{i}(q)
&= E_{i}+\left[1-\exp\!\left(-c_{\mathrm{disp}}\,q^{d_{\mathrm{disp}}}\right)\right]
\,\mathrm{Ry}\,q^{2}, \\[4pt]
\gamma_{k}(q)
&= \gamma_{k}+b_{1}\,(\mathrm{Ry}\,q)+b_{2}\,(\mathrm{Ry}\,q^{2}),
\qquad k\in\{j,i\}.
\end{align}
\end{subequations}

The excitation oscillator strengths $f_j(q)$ are prescribed explicitly for each band. The ionization strengths $f_i(q)$ are then rescaled so that the total oscillator strength remains fixed at each $q$. This enforces conservation of the total spectral weight, namely the oscillator-strength sum rule, or $f$-sum rule: the total oscillator strength (spectral weight) is conserved as the momentum transfer varies. The ionization energies $E_i(q)$ are dispersed through a saturating shift toward the free-particle scaling $\mathrm{Ry}\,q^2$, while the widths $\gamma_k(q)$ for both excitation and ionization channels acquire additional momentum-dependent broadening.
This dispersion scheme reduces to the optical limit as $q\to 0$, is consistent at small $q$ with the dielectric response of a homogeneous electron gas in the random-phase approximation (RPA) \citep{pines1952collective} and in its finite-damping Mermin extension \citep{mermin1970lindhard}, and approaches the large-$q$ impulse regime, in which the characteristic energy transfer scales as $E\sim \mathrm{Ry}\,q^2$ \citep{emfietzoglou2007consistent,emfietzoglou2017monte}. In the RPA, electrons of the target medium respond collectively to the perturbing particle through a self-consistent screening field, while correlations beyond that mean-field response are neglected. The Mermin extension retains this collective description, but introduces a finite relaxation time so that excitations may broaden and decay without violating local charge conservation.

The imaginary dielectric response, $\varepsilon_2(E,q)$, is then constructed by evaluating the optical-limit Drude kernels with the $q$-dependent parameters of Equation~\ref{eq:ecn_dispersion}. In the transport implementation, each excitation and ionization channel is imposed only above its channel-specific threshold, as specified by the corresponding $B_{\mathrm{th}}$ values. The per-excitation dispersion triplets $(a_j,b_j,c_j)$ and the global coefficients $(c_{\mathrm{disp}},d_{\mathrm{disp}},b_1,b_2)$ are listed in Appendix~\ref{app:excit_ion}.

\subsection{From dielectric properties to differential cross-sections}\label{sec:excitation_ionization_cs}

When the imaginary part of the dielectric function ($\varepsilon_2$) is written as a sum of contributions from discrete excitation bands and ionization shells (Equation~\ref{eq:epsilon2}), the ELF inherits the same additive structure,

\begin{equation}
\begin{aligned}
\mathrm{ELF}(E,q)
&=\sum_{j=1}^{N_{\mathrm{exc}}}\mathrm{ELF}^{(\mathrm{exc})}_j(E,q)\\
&\quad+\sum_{i=1}^{N_{\mathrm{ion}}}\mathrm{ELF}^{(\mathrm{ion})}_i(E,q).
\end{aligned}
\end{equation}

\begin{equation}
\begin{aligned}
\mathrm{ELF}^{(k)}(E,q)
&=\frac{\varepsilon_2^{(k)}(E,q)}
{\varepsilon_1^{2}(E,q)+\varepsilon_2^{2}(E,q)}\,,
\end{aligned}
\end{equation}

with $k$ denoting a specific excitation band or ionization shell. This partitioning enables channel-resolved cross-sections while preserving the full screening encoded in the total denominator.
Within the plane-wave Born approximation (PWBA), the incident and scattered electrons are treated as plane waves, and the interaction with the medium is evaluated to first order in the perturbing potential. For an amorphous condensed medium, the resulting double-differential inelastic cross-section per molecule takes the form
\begin{equation}
\frac{\mathrm{d}^2\sigma(E,q;T)}{\mathrm{d}E\,\mathrm{d}q}
= \frac{1}{\pi a_0 N\,T}\,\frac{1}{q}\,\mathrm{ELF}(E,q),
\end{equation}
where $N$ is the molecular number density. Integrating over the kinematically allowed momentum transfers yields the energy-transfer \textbf{inelastic} differential cross-section (DCS)
\begin{equation}
\frac{\mathrm{d}\sigma(E;T)}{\mathrm{d}E}
= \frac{1}{\pi a_0 N\,T}
\int_{q_{\min}(E,T)}^{q_{\max}(E,T)} \frac{\mathrm{d}q}{q}\,\mathrm{ELF}(E,q).
\end{equation}
Channel-resolved inelastic DCS are obtained by replacing $\mathrm{ELF}$ with $\mathrm{ELF}^{(\mathrm{exc})}_j$ or $\mathrm{ELF}^{(\mathrm{ion})}_i$ in the integrand \citep{emfietzoglou2017monte}.

For a nonrelativistic projectile and scalar $q$, the integration bounds follow directly from energy and momentum conservation,
\begin{equation}
\begin{aligned}
q_{\min}(E,T)=\sqrt{2m}\Big(\sqrt{T}-\sqrt{T-E}\Big),\\
\qquad
q_{\max}(E,T)=\sqrt{2m}\Big(\sqrt{T}+\sqrt{T-E}\Big),
\end{aligned}
\end{equation}
with $m$ the electron mass (in the same unit system as $T$ and $E$).

Finally, the total cross-section (TCS) for a given \textbf{inelastic} channel is obtained by integrating its DCS over the appropriate energy-transfer window,
\begin{equation}
\sigma^{(k)}(T)
=\int_{E_{\min}^{(k)}}^{E_{\max}^{(k)}}
\frac{\mathrm{d}\sigma^{(k)}(E;T)}{\mathrm{d}E}\,\mathrm{d}E.
\end{equation}

\begin{equation}
\begin{aligned}
E_{\min}^{(\mathrm{exc},j)} &= B_{\rm min}, &\qquad E_{\max}^{(\mathrm{exc},j)} &= T,\\
E_{\min}^{(\mathrm{ion},i)} &= B_{\mathrm{th},i}, &\qquad E_{\max}^{(\mathrm{ion},i)} &= \frac{T+B_{\mathrm{th},i}}{2},\\
E_{\min}^{(K)}              &= B_{\mathrm{th},K}, &\qquad E_{\max}^{(K)}              &= \frac{T+B_{\mathrm{th},K}}{2}. 
\end{aligned}
\end{equation}

where $B_{\rm min}$ is the threshold energy of electronic excitations (equal to the band-gap energy), $B_{\mathrm{th},i}$ is the binding-energy threshold of the $i^{\rm th}$ ionization shell, and $B_{\mathrm{th},K}$ is the oxygen K-shell threshold.
Finally, to generate the angular dependence, we evaluate the finite-$q$ kernel and map the momentum transfer to the scattering angle using relativistic kinematics. For a given incident kinetic energy $T$ and sampled energy transfer $E$, the post-collision kinetic energy is $T' = T - E$, and
\begin{equation}
\cos\theta \;=\; \frac{p^2(T) + p^2(T') - q^2}{2\,p(T)\,p(T')}\,,
\end{equation}
where $p(T)$ and $p(T')$ are the magnitudes of the incident and outgoing electron momenta, respectively, defined by the relativistic energy--momentum relation
\begin{equation}
p(T)c \;=\; \sqrt{T\!\left(T+2m_ec^2\right)}\,.
\end{equation}
The angle-differential distribution follows from the change of variables $q \mapsto \cos\theta$, with Jacobian
\begin{equation}
\left|\frac{\mathrm{d}q}{\mathrm{d}\cos\theta}\right|
\;=\;
\frac{p(T)\,p(T')}{q}\,,
\end{equation}
so that
\begin{align}
\frac{\mathrm{d}\sigma^{(k)}(E;T)}{\mathrm{d}\Omega}
&=
\frac{1}{2\pi}\,
\frac{\mathrm{d}\sigma^{(k)}(E;T)}{\mathrm{d}\cos\theta}
\nonumber\\
&=
\frac{1}{2\pi}\,
\frac{\mathrm{d}\sigma^{(k)}(E,q;T)}{\mathrm{d}E\,\mathrm{d}q}\,
\left|\frac{\mathrm{d}q}{\mathrm{d}\cos\theta}\right|
\nonumber\\
&=
\frac{1}{2\pi}\,
\frac{\mathrm{d}\sigma^{(k)}(E,q;T)}{\mathrm{d}E\,\mathrm{d}q}\,
\frac{p(T)\,p(T')}{q}\,.
\end{align}

\begin{figure*}[t!]
\centering
\rotatebox[origin=c]{0}{\includegraphics[scale = 0.45]{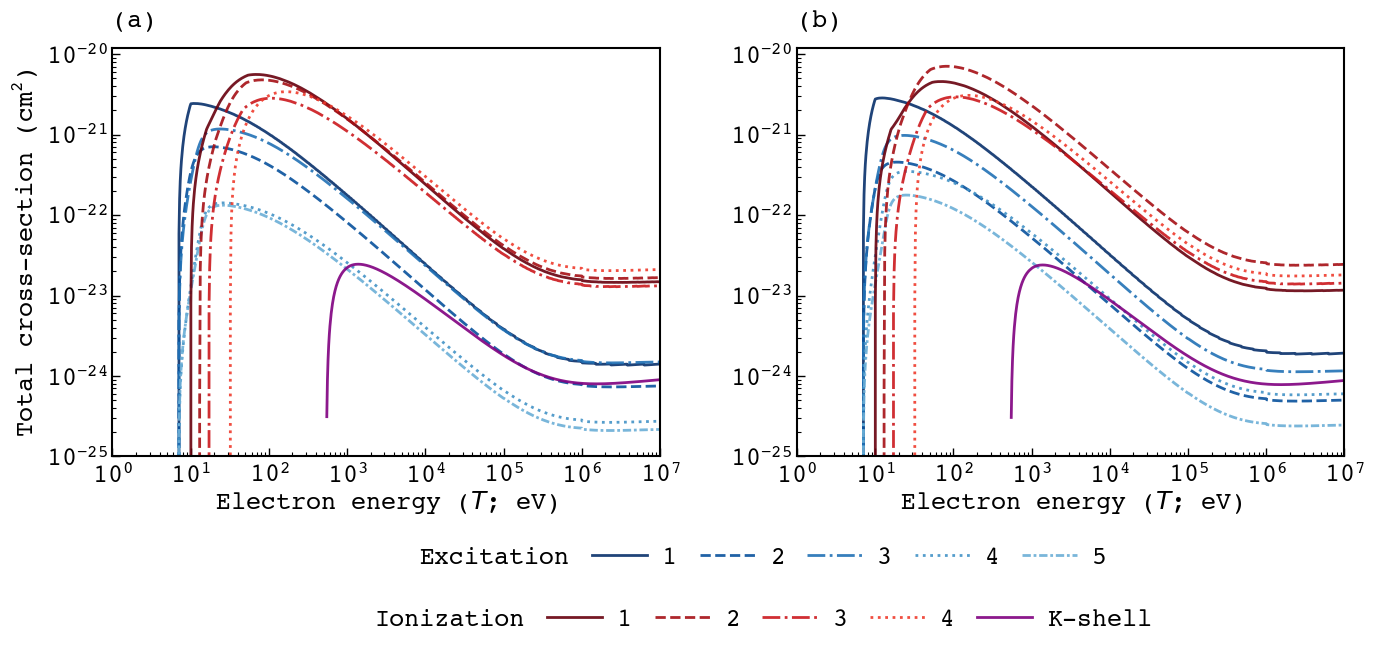}}
\caption{Total per-channel cross-sections for five ionization (red curves; O K-shell in purple) and five excitation (blue curves) channels in (\textbf{a}) amorphous and (\textbf{b}) hexagonal ice after relativistic correction.
\textbf{Note:} The plotted cross-sections are molecular cross-sections computed for ice density of 1~g~cm$^{-3}$. In Geant4-IcyMoons, bulk density effects are accounted for by using molecular number density to scale interaction rates at runtime.
}
 \label{fig:total_cs_exc_ion}
\end{figure*}

At higher incident energies, the underlying inelastic DCS must be evaluated with relativistic kinematics and, where applicable, transverse and density-effect contributions \citep{kyriakou2025extension}; in our implementation, these corrections modify the integrand $\mathrm{d}\sigma^{(k)}(E;T)/\mathrm{d}E$ but do not alter the energy-transfer bounds above. A complete description of the relativistic formulation and the regime logic used to activate each correction is provided in Appendix~\ref{app:excit_ion}. Figure~\ref{fig:total_cs_exc_ion}a shows the corrected total cross-sections of all excitation and ionization channels for amorphous and hexagonal ice.

\section{Benchmarking}\label{sec:benchmarking}

We benchmark three standard low-order energy-loss observables: the electronic stopping power, the $W$-value (mean energy expended per ion pair), and the inelastic mean free path. The stopping power is evaluated deterministically from the inelastic interaction model, whereas the $W$-value is obtained from dedicated Geant4-IcyMoons track-structure simulations that explicitly follow ionization cascades and count the total number of ionizations.

\subsection{Stopping power}

The electronic stopping power, $S(T)\equiv-\mathrm{d}T/\mathrm{d}x$, is evaluated as the first energy-transfer moment of the inelastic interaction model. For a homogeneous medium of molecular number density $N$,
\begin{equation}
S(T)=N\sum_{c}\int E\,\frac{\mathrm{d}\sigma_c(T,E)}{\mathrm{d}E}\,\mathrm{d}E,
\end{equation}
with discrete sums for channels represented as level-resolved total cross sections. Excitation losses are computed as level-energy sums weighted by the corresponding total cross sections. Ionization losses are obtained by shell-wise integration of $E\,\mathrm{d}\sigma/\mathrm{d}E$ over
\begin{equation}
E \in \left[B_{\mathrm{th},i},\frac{T+B_{\mathrm{th},i}}{2}\right],
\end{equation}
for shell threshold $B_{\mathrm{th},i}$. Attachment is treated as terminal, contributing $N\,T\,\sigma_{\mathrm{att}}(T)$. The totals shown in Figure~\ref{fig:stopping_power}a include all baseline inelastic channels.

At low energies, the liquid-water and ice treatments diverge. The stopping-power curve of the former is defined only down to about $8~\mathrm{eV}$ through the excitation model, whereas electron transport itself terminates at $10~\mathrm{eV}$ and is handed off to solvation, that is, localization of the sub-excitation electron in the liquid. In Geant4-IcyMoons, by contrast, transport in ice is followed down to $2~\mathrm{eV}$. In this regime, stopping power is governed mainly by vibrational excitation, while dissociative attachment contributes only over a narrow resonance interval and acts primarily as a terminal sink. At higher energies, the shape of the ice stopping-power curve follows the main structure of the modeled ELF. Its initial rise is set by the opening of the valence-excitation manifold above $7~\mathrm{eV}$ (see Section~\ref{sec:excitation_ionization}), which in Geant4-IcyMoons is described by five bands centered between about $9$ and $15~\mathrm{eV}$ (see Figure~\ref{fig:total_cs_exc_ion}). The broad maximum at $\mathcal{O}(100~\mathrm{eV})$ is instead driven mainly by the broader valence-ionization structure between about $15$ and $40~\mathrm{eV}$ (see Figure~\ref{fig:total_cs_exc_ion}). In the finite-$q$ extension, these ionization bands broaden and shift to higher energy transfer; thus, their contribution remains substantial away from the optical limit. Because stopping power is the first moment of the inelastic differential cross section, these higher-energy losses weigh more strongly and therefore set the location of the maximum. The oxygen K-shell enters only above $\sim540~\mathrm{eV}$ and mainly affects the high-energy tail. The larger separation of liquid water from the two ice phases reflects differences in both molecular number density and electronic inelastic response. By contrast, amorphous and hexagonal ice exhibit closely related solid-phase dielectric responses and share the remaining low-energy channels.

\subsection{$W$-value}

The $W$-value is obtained directly from Geant4-IcyMoons event statistics, following the standard cascade-based definition used in Geant4-DNA. For all monoenergetic primaries at incident kinetic energy $T$, we score the total number of electron-impact ionization interactions, $N_{\mathrm{ion}}^{(m)}$,
produced by primary $m$ (including ionizations produced by the full secondary cascade). The reported $W$-value is then
\begin{equation}
W(T)=\frac{T}{\langle N_{\mathrm{ion}}\rangle},
\qquad
\langle N_{\mathrm{ion}}\rangle=\frac{1}{N_{\mathrm{evt}}}\sum_{m=1}^{N_{\mathrm{evt}}}N_{\mathrm{ion}}^{(m)}.
\end{equation}

Figure~\ref{fig:stopping_power}b shows the numerically evaluated $W(T)$ for liquid water, amorphous ice, and hexagonal ice. The $W$-value complements the stopping power by measuring how efficiently deposited energy is converted into ion pairs, whereas the latter measures energy lost per unit path length; equivalently, $W$ reflects the ratio of total energy loss to the ionization yield, set by the ionization cross section and secondary production. At low energies, all three targets show broadly similar behavior: for small incident energies near the ionization threshold, the number of energetically allowed ionization channels is small, thus a larger fraction of the incident energy is spent in sub-ionizing losses, and $W$ rises rapidly. At higher energies, $W$ flattens. In this regime, the two ice phases lie slightly above the liquid water line, indicating a modestly lower ionization efficiency in ice. The much smaller separation between amorphous and hexagonal ice, consistent with the stopping-power behavior (see Figure~\ref{fig:stopping_power}a), reflects their closely related valence-ionization structure.

\subsection{Inelastic mean free path}

\begin{figure}[t!]
\centering
\includegraphics[scale=0.32]{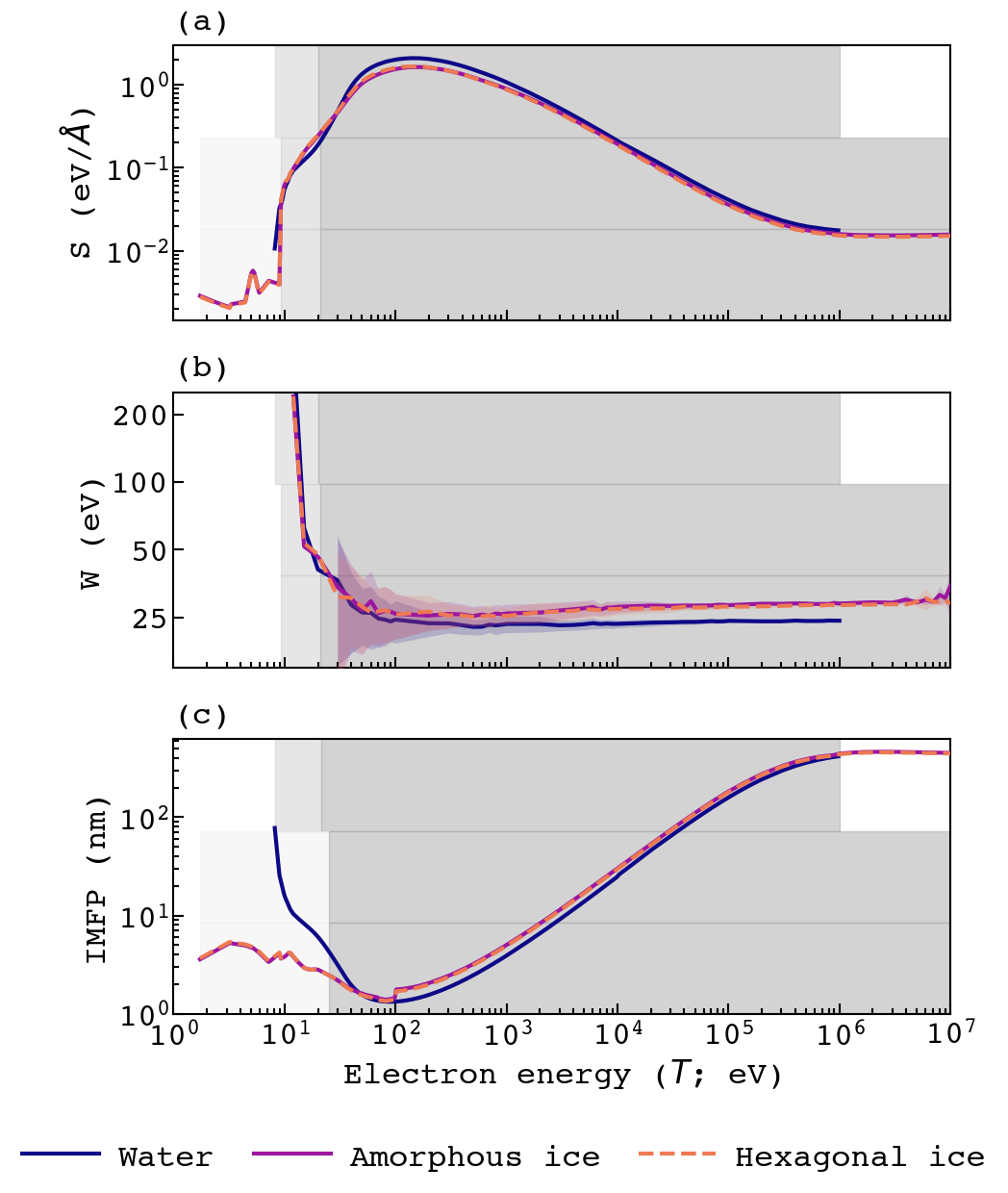}
\caption{Geant4-IcyMoons benchmarks. (\textbf{a}) Electron stopping power in liquid water (\texttt{G4EmDNAPhysics\_option4}) compared with amorphous and hexagonal ice; (\textbf{b}) Corresponding W-value (mean energy expended per ion pair), assuming bulk mass densities of 0.94 g~cm$^{-3}$ and 0.917 g~cm$^{-3}$ for amorphous and hexagonal ice, respectively \citep{petrenko1999physics, michaud2003cross, amato2026molecular}, with 1$\sigma$ uncertainties shown in colored bands; (\textbf{c}) inelastic mean free path (IMFP) derived from all inelastic channels. The shaded background indicates the dominant inelastic channel in each energy range: vibrational excitation + dissociative (de-)attachment (light), excitation (mid), and ionization (dark). The upper third shows water dominance, the middle third amorphous ice, and the lower third hexagonal ice. Sharp transitions in the bands indicate only a change in the channel whose contribution is largest.}
\label{fig:stopping_power}
\end{figure}

Figure~\ref{fig:stopping_power}c shows the inelastic mean free path (IMFP), $\lambda_{\rm inel}=(N\sigma_{\rm inel})^{-1}$, for liquid water and both ice phases. At low energies, electrons in ice exhibit a shorter mean free path than in liquid water. In the present implementation, this mainly reflects the strong contribution of vibrational excitation and dissociative attachment to the low-energy inelastic cross section in ice. This does not contradict the stopping-power behavior in Figure~\ref{fig:stopping_power}a: the IMFP is controlled by how frequently inelastic events occur, whereas the stopping power weights those same events by their transferred energy. Vibrational excitation, therefore, shortens the IMFP efficiently, while contributing more modestly to the stopping power as each event carries only a small energy loss. Dissociative attachment, by contrast, acts over a narrow energy interval and terminates the cascade upon capture. In the default liquid-water comparison, the inelastic response begins with the electronic channels. For energies $<10~\mathrm{eV}$, the comparison between liquid water and ice should therefore be interpreted as model-dependent rather than as a strict material comparison.

At $100~\mathrm{eV}$, the IMFP increases abruptly in the two ice curves. This is a direct consequence of the present channel construction: the vibrational contribution vanishes at the upper limit of the tabulated low-energy cross sections. Once that channel is removed, the total inelastic cross section decreases, and the IMFP correspondingly lengthens. Above this transition, the ice IMFP is mainly controlled by electronic channels. The two ice phases remain close throughout, consistent with their closely related phase-resolved electronic response, while the remaining low-energy channels are treated identically to first order.

\section{Out-of-model handling}\label{sec:outofrange}

Outside the range for which Geant4-IcyMoons is defined ($T \in [1,10^7]~\mathrm{eV}$), particle transport is closed through explicit boundary rules. This is consistent with the hybrid construction adopted here, in which the combined interaction model is intended to operate only within its validated domain. Below, we discuss all cases in which such closures apply.

\subsection{Low-energy termination} \label{sec:outofrange_low}

At the low-energy boundary, electrons entering the sub-excitation regime are not transported further. When an electron reaches $T < 2~\mathrm{eV}$, we terminate its track and deposit its remaining kinetic energy locally. 
In liquid water, the analogous endpoint is commonly treated as solvation (hydrated-electron formation) followed by chemical evolution \citep{incerti2010comparison}. In solid ice, the corresponding endpoint is better interpreted as localization into trapped excess-electron states, but trapped-electron dynamics (detrapping, hopping, recombination) are outside the scope of the present physical stage \citep[e.g.,][]{bhattacharya2014excess}. Accordingly, the $2~\mathrm{eV}$ cutoff should be considered a transport termination rule that marks the handoff point, rather than as an assertion about a unique microscopic timescale.

\subsection{High-energy handoff} \label{sec:outofrange_high}

At the upper boundary of the Geant4-IcyMoons domain, electron transport may be continued using the different available suites of standard Geant4 electromagnetic (EM) physics.
In this regime, angular deflection is handled by the multiple Coulomb scattering and single Coulomb scattering formalisms \citep{urban2006model}, 
while the energy-loss rate is governed by Bethe-type ionization \citep{agostinelli2003geant4, allison2016recent}. 
This handoff avoids extrapolation of the ice-specific low-energy interaction kernels into a regime where independent-scatterer assumptions become appropriate \citep{apostolakis2010validation}.

\subsection{Bremsstrahlung} \label{sec:outofrange_brems}

Bremsstrahlung, namely photon emission by electrons in the Coulomb field of nuclei or atomic electrons, is not modeled explicitly in Geant4-IcyMoons. Instead, for $T \ge 1\,\mathrm{MeV}$ we defer to the standard Geant4 electromagnetic bremsstrahlung process \citep{allison2016recent}, implemented by interpolation of tabulated differential cross sections \citep{seltzer1985bremsstrahlung,seltzer1986bremsstrahlung}. In this treatment, photons above the production threshold are generated explicitly, whereas softer emission is absorbed into the continuous energy loss. We place the handoff at $1\,\mathrm{MeV}$ because in water-like low-$Z$ media radiative losses remain subdominant to collisional losses far below the electron critical energy of liquid water, $E_c \approx 78\,\mathrm{MeV}$ \citep{navas2024review}. Below this threshold, bremsstrahlung has little effect on the total electron energy-loss budget, so the ice-specific low-energy treatment is retained without an explicit radiative channel.

\section{Use case: electron bombardment of Europa}\label{sec:Europa}

\begin{figure}
\centering
\rotatebox[origin=c]{0}{\includegraphics[scale = 0.32]{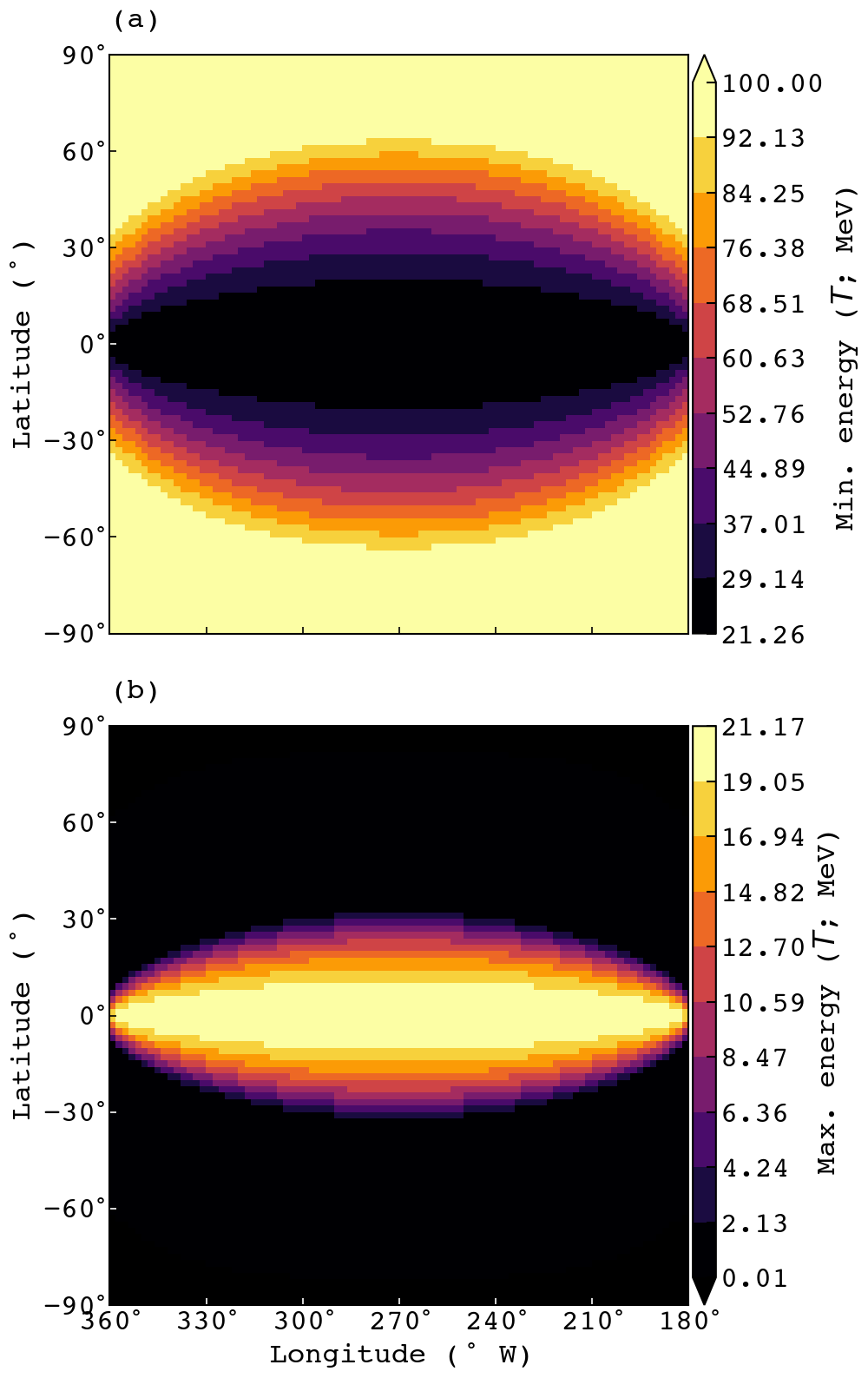}}
  \caption{Electron-access energy maps on Europa, following \citet{nordheim2018preservation}. At each surface coordinate, these maps indicate the range of incident electron energies able to reach that location. (\textbf{a}) Lower bound of that accessible energy range on the leading hemisphere; (\textbf{b}) The corresponding upper bound on the trailing hemisphere.}
     \label{fig:europa_electrons_map}
\end{figure}

\begin{figure*}
\centering
\rotatebox[origin=c]{0}{\includegraphics[scale = 0.32]{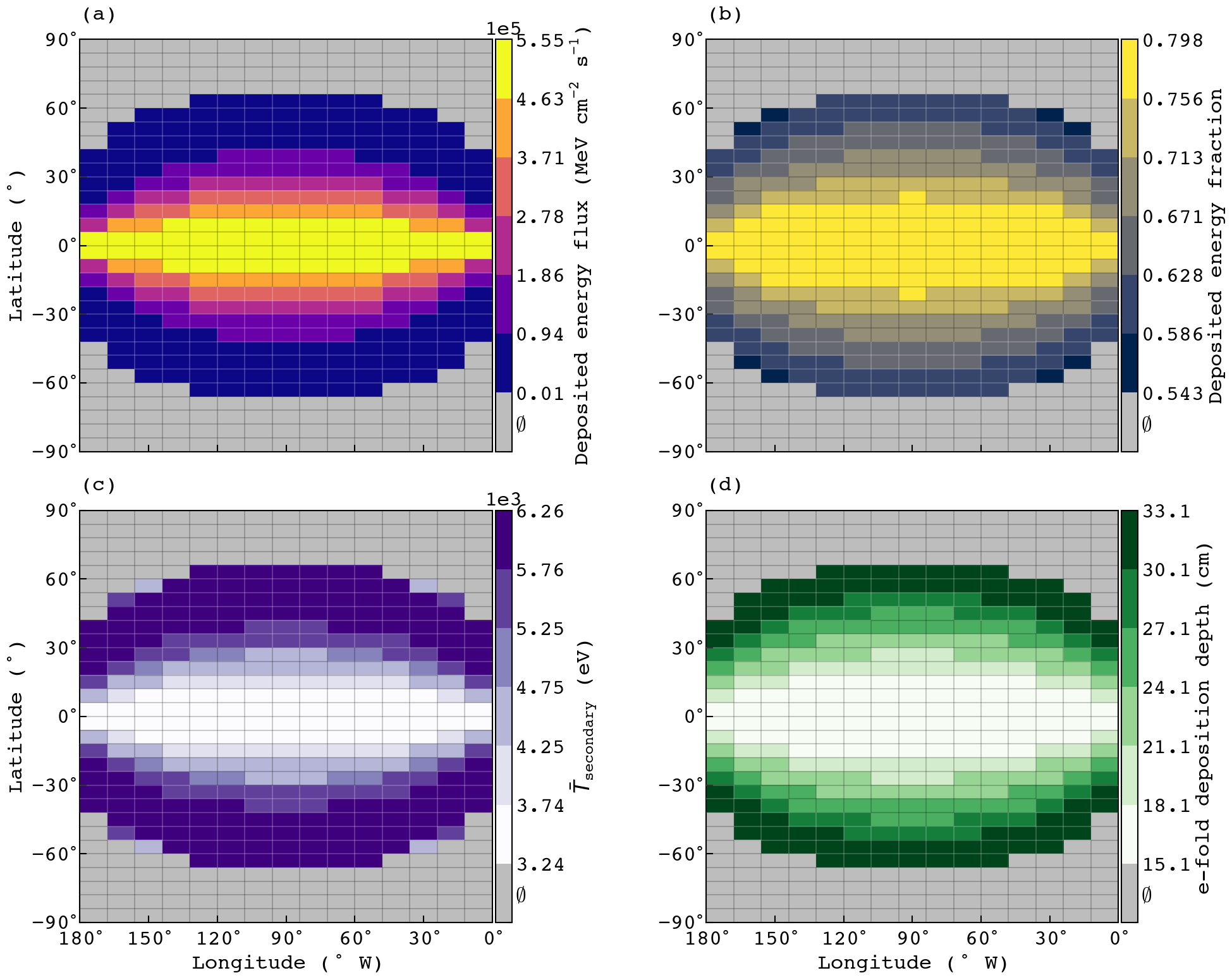}}
  \caption{Four diagnostics of electron bombardment on Europa's leading hemisphere: (\textbf{a}) total deposited-energy flux, (\textbf{b}) deposited-energy fraction, (\textbf{c}) mean secondary-electron energy, and (\textbf{d}) e-folding ($\approx 63\%$) deposition depth. Cells with no electron flux in the relevant energy range are shown in gray.}
     \label{fig:europa_4p_leading}
\end{figure*}

\begin{figure*}
\centering
\rotatebox[origin=c]{0}{\includegraphics[scale = 0.32]{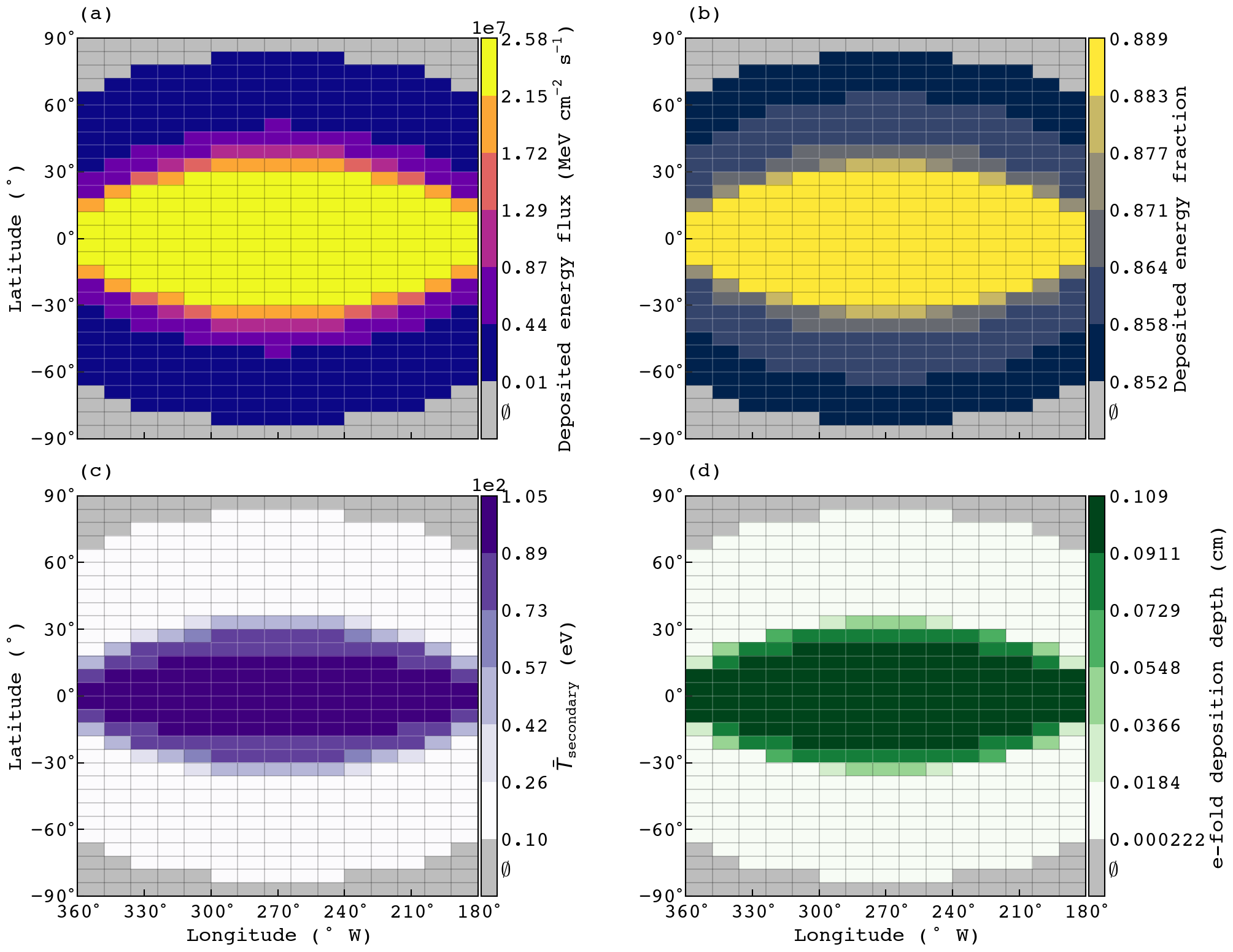}}
  \caption{Four diagnostics of electron bombardment on Europa's trailing hemisphere, formatted similarly to Figure~\ref{fig:europa_4p_leading}.}
     \label{fig:europa_4p_trailing}
\end{figure*}

\begin{figure}
\centering
\rotatebox[origin=c]{0}{\includegraphics[scale = 0.3]{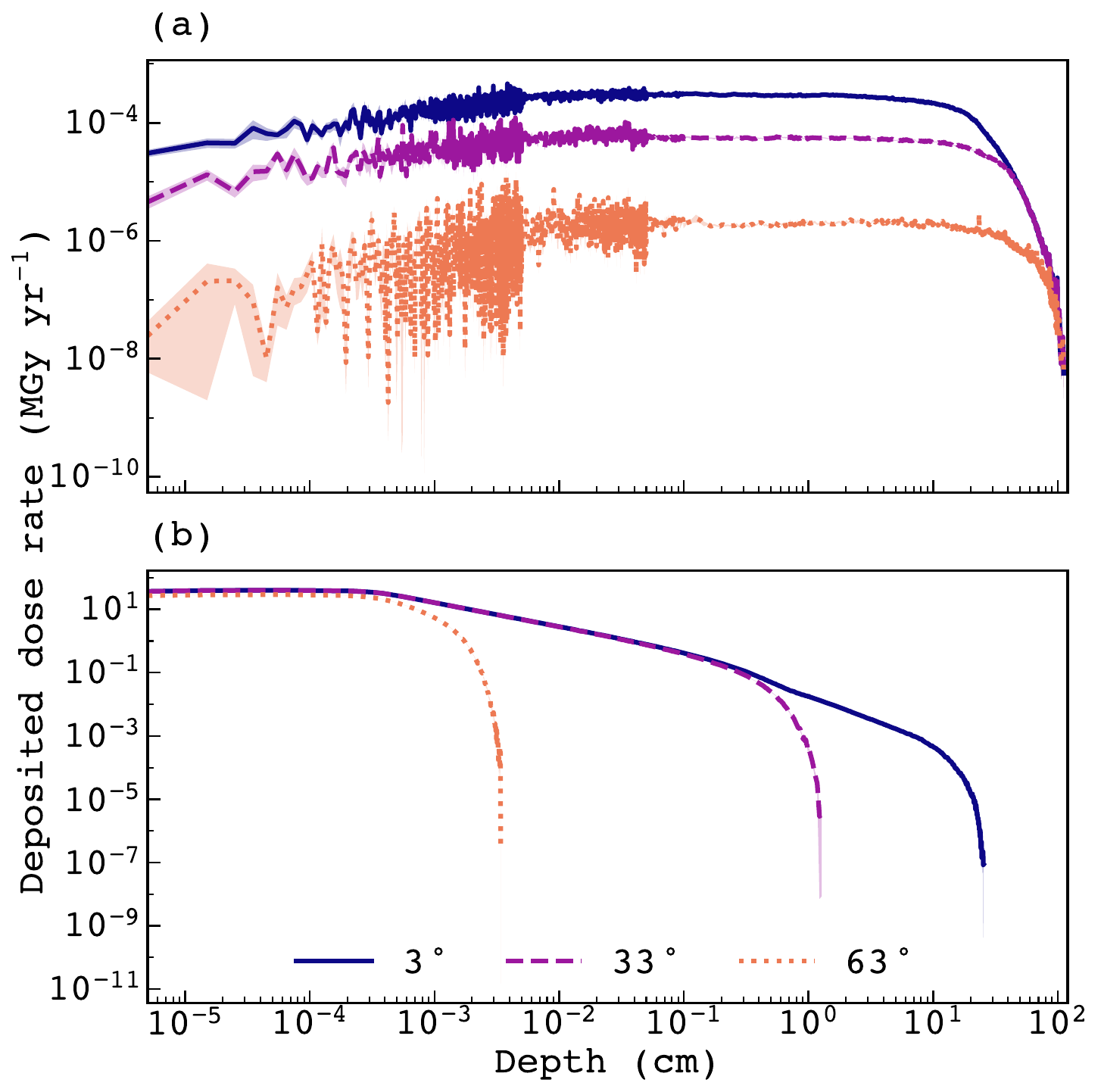}}
\caption{
Depth-resolved deposited-dose rates from electron bombardment in Europa's near-surface ice of density 0.5 g cm$^{-3}$, shown for three nominal latitudes centered on the prime meridians of the (\textbf{a}) leading and (\textbf{b}) trailing hemispheres. Dose rates are plotted as a function of depth for latitudes $3^\circ$, $33^\circ$, and $63^\circ$. Colored bands indicate the $1\sigma$ event-level spread in deposited dose propagated across the simulated incident-electron energy spectrum.}
\label{fig:europa_MGy}
\end{figure}

Irradiation by magnetospheric plasma is a dominant agent shaping Europa's near-surface ice \citep{paranicas2001electron, nordheim2022magnetospheric}. In particular, electrons with kinetic energies of roughly 0.1--100~MeV govern the spatial distribution of deposited energy, drive radiolytic chemistry, and modulate the balance between amorphization and recrystallization of water ice \citep{strazzulla1992ion,loeffler2020possible}. These processes imprint themselves on Europa’s surface through hemispheric asymmetries in ice phase \citep{cartwright2025jwst, Yoffe_2026}, chemical composition, and the accumulation of radiation products \citep{wu2024europa}. Most notable such products are sulfate-rich assemblages on the trailing hemisphere, which are widely interpreted as the outcome of sustained radiolysis \citep{carlson1999sulfuric, carlson2005distribution}. The same radiation environment governs the production, modification, and destruction of embedded molecular species, constraining the longevity of chemically informative compounds and potential biosignatures exposed at or near the surface \citep{yoffe2025fluorescent}.

A primary application of Geant4-IcyMoons is the simulation of this magnetospheric electron bombardment. Electrons sampled from prescribed energy spectra, obtained from spacecraft measurements, such as \textit{Voyager} and \textit{Galileo} \citep{paranicas2001electron}, are injected into amorphous water ice and transported through the condensed medium to compute depth-dependent energy deposition and secondary electron production. The resulting radiation fields provide the physical basis for quantifying radiation-driven ice evolution, radiolytic chemistry, and the degradation of embedded molecular species under Europa-relevant conditions \citep[e.g.,][]{nordheim2018preservation, yoffe2025fluorescent}. This electron-focused treatment will later be complemented by the inclusion of dominant magnetospheric ions, including protons and heavier oxygen and sulfur species \citep{nordheim2022magnetospheric}. The resulting energy-deposition profiles encode the full incident electron spectrum. The hemispheric and latitudinal structure of the modeled surface energy deposition \citep[following calculations presented in][]{nordheim2018preservation}, and reflecting the spatial organization of magnetospheric electron fluxes, is shown in Figure~\ref{fig:europa_electrons_map}. Four diagnostics of electron bombardment of Europa's surface are presented for the leading and trailing hemispheres in Figure~\ref{fig:europa_4p_leading} and Figure~\ref{fig:europa_4p_trailing}, respectively. Depth-resolved deposited-dose profiles for three nominal latitudes centered on the prime meridians of both hemispheres are shown in Figure~\ref{fig:europa_MGy}.

The modeled energy-deposition fields reveal a pronounced hemispheric asymmetry in both the depth and intensity of electron-driven processing. On the trailing hemisphere, where Europa is exposed to a high flux of lower-energy electrons, deposition is concentrated within a shallow, optically active veneer, with characteristic penetration depths typically $\lesssim 0.1$ cm. On the leading hemisphere, by contrast, the incident population is less intense but more energetic, driving processing to substantially greater depths, reaching tens of centimeters. Their deposition efficiency is also lower, since a larger fraction of the incident energy is carried to depth or lost radiatively through bremsstrahlung rather than deposited locally near the surface (see Appendix~\ref{app:Europa}). In both hemispheres, however, most of the deposited energy remains concentrated at equatorial and low latitudes, tracing the lens-like spatial organization of Europa's magnetospheric electron bombardment \citep{paranicas2001electron}. In the near-equatorial regions, the deposited dose rates therefore differ by several orders of magnitude between the two hemispheres, underscoring that Europa's radiation environment is asymmetric not only in penetration depth, but also in the efficiency with which energy is delivered to near-surface layers.

This asymmetry is likely relevant to Europa’s long-recognized compositional dichotomy. The trailing hemisphere, together with adjacent low-albedo terrains, is associated with hydrated sulfuric acid and other sulfur-bearing radiolytic products \citep{carlson1999sulfuric,carlson2005distribution}. Sulfur-ion access is itself strongly structured: lower-energy sulfur ions preferentially reach the trailing hemisphere overall, but with reduced access across parts of the equatorial trailing hemisphere, whereas the highest-energy sulfur ions access the surface more nearly uniformly \citep{nordheim2022magnetospheric}. Superimposed on this, magnetospheric electrons provide a strongly asymmetric energy input, with especially shallow and intense deposition on the trailing hemisphere (see Figure~\ref{fig:europa_4p_leading} and Figure~\ref{fig:europa_4p_trailing}). The observed hemispheric contrast, therefore, likely reflects the combined effects of spatially structured sulfur implantation and electron-driven radiolytic at the uppermost surface layer, with the lens-like trailing-hemisphere morphology likely tracking the latter more directly.

In that context, the shallow and intense trailing-hemisphere energy deposition by electrons may be especially effective at modifying the optically active surface veneer, thereby helping sustain the sulfuric-acid-rich, low-albedo pattern \citep{mergny2025bond}. The leading hemisphere, although exposed to more penetrating electrons, distributes a larger fraction of its electron-driven processing to greater depths and therefore does not need to develop an equally strong surficial radiolytic signature \citep{cartwright2025jwst,Yoffe_2026}. Taken together, these results support an interpretation in which, once sulfur-bearing precursors are emplaced, electron-driven radiolysis is a primary agent shaping and maintaining the hemispheric distribution of sulfuric acid and associated dark material on Europa's surface.

\section{Conclusion and outlook} \label{conclusions}

This first implementation of Geant4-IcyMoons establishes a physically grounded description of electron transport in condensed water ice and demonstrates its relevance to Europa's near-surface radiation environment. By resolving how incident electrons deposit energy as a function of depth, latitude, and hemisphere, the model provides the missing transport layer between magnetospheric forcing and the physical and chemical evolution of irradiated ice. On Europa, this link is essential: the observable surface is not a passive record of composition, but a dynamically processed interface in which bombardment, transport, and local ice properties jointly regulate what is produced, retained, and ultimately expressed as surface observables.

The Geant4 classes implementing electron interactions in water ice are publicly available as part of Geant4-IcyMoons \citep{icymoonsrelease}. In the future, we intend to extend this framework beyond electron transport alone, toward a unified description of Europa's near-surface processing environment. This includes the dominant magnetospheric ions — particularly protons and oxygen- and sulfur-bearing species \citep{nordheim2022magnetospheric}, which both implant chemically active material and deposit energy through track structures distinct from those of electrons \citep{yoffe2025fluorescent}, thereby coupling precursor delivery and radiolytic forcing within the same transport framework. It also requires coupling the resulting energy-deposition fields to radiolytic chemistry in cryogenic ices, where product yields, back-reactions, and net chemical evolution depend on temperature, composition, mixing state, and dose history.

In addition to altering near-surface chemical composition, the irradiation environment also acts together with sputtering \citep{vorburger2018europa}, sintering \citep{molaro2019microstructural, mergny2024lunaicy}, amorphization and recrystallization of water-ice, porosity evolution, and regolith gardening \citep{costello2021impact} to determine the lifetimes of various materials embedded in the near-surface regolith and the textures of the near-surface regolith. On Europa, such coupling is essential: radiolytic lifetimes are short compared to geologic timescales, sputtering can counteract net chemical buildup, and the observable surface records the competition among irradiation, transport, and microphysical reworking rather than composition alone \citep{ligier2016vlt, thelen2024subsurface, cartwright2025jwst, Yoffe_2026}.

Ultimately, our goal is to model a self-consistent organic and inorganic surface cycle encompassing implantation, radiolytic transformation, sputter loss, burial, and re-exposure. A framework of this kind would connect incident particle populations to observables such as radiolytic species abundances, oxidant production, sputtering yields, sintering and phase-evolution timescales, and the survivability of carbon-bearing, potentially biosignature-relevant compounds. In this sense, Geant4-IcyMoons serves as a physical backbone for future coupled descriptions of radiation processing in astrophysical ices across planetary, satellite, and small-body environments.

\begin{acknowledgments}
We are thankful to Guy Ron, Shigeo Tanuma, and Ruth Signorell for useful discussions, and to Jonathan Lunine for reviewing this manuscript prior to submission.
We are also thankful to the Helen Kimmel Center for Planetary Science at the Weizmann Institute of Science for its assistance in funding this study. Additionally, we are thankful to the anonymous reviewer, whose thoughtful comments strengthened the clarity and impact of this work.
\end{acknowledgments}

\FloatBarrier
\bibliography{sample}{}

@string{jgrp = {Journal of Geophysical Research: Planets}}

@string{nata = {Nature Astronomy}}

@string{plscij = {Planetary Science Journal}}

@string{po = {Progress in Oceanography}}

@article{nordheim2018preservation,
  title = {Preservation of potential biosignatures in the shallow subsurface of {E}uropa},
  author = {Nordheim, T. A. and Hand, K. P. and Paranicas, C.},
  journal = {Nature Astronomy},
  volume = {2},
  number = {8},
  pages = {673--679},
  year = {2018},
  publisher = {Nature Publishing Group}
}

@article{henning2013chemistry,
  title={Chemistry in protoplanetary disks},
  author = {Henning, T. and Semenov, D.},
  journal={Chemical Reviews},
  volume={113},
  number={12},
  pages={9016--9042},
  year={2013},
  publisher={ACS Publications}
}

@article{he2022refractive,
  title = {Refractive Index and Extinction Coefficient of Vapor-deposited Water Ice in the UV--vis Range},
  author = {He, J. and Diamant, S. J. M. and Wang, S. and Yu, H. and Rocha, W. R. M. and Rachid, M. and Linnartz, H.},
  journal = {Astrophysical Journal},
  volume = {925},
  number = {2},
  pages = {179},
  year = {2022},
  publisher = {IOP Publishing}
}

@article{agostinelli2003geant4,
  title = {Geant4—a simulation toolkit},
  author = {Agostinelli, S. and Allison, J. and Amako, K. and Apostolakis, J. and Araujo, H. and Arce, P. and Asai, M. and Axen, D. and Banerjee, S. and Barrand, G. J. N. I. and others},
  journal = {Nuclear Instruments and Methods in Physics Research Section A: Accelerators, Spectrometers, Detectors and Associated Equipment},
  volume = {506},
  number = {3},
  pages = {250--303},
  year = {2003},
  publisher = {Elsevier}
}

@article{paranicas2001electron,
  title = {Electron Bombardment of {E}uropa},
  author = {Paranicas, C. and Carlson, R. W. and Johnson, R. E.},
  journal = {Geophysical Research Letters},
  volume = {28},
  number = {4},
  pages = {673--676},
  year = {2001},
  publisher = {Wiley Online Library}
}

@article{nordheim2022magnetospheric,
  title = {Magnetospheric Ion Bombardment of {E}uropa’s Surface},
  author = {Nordheim, T. A. and Regoli, L. H. and Harris, C. D. K. and Paranicas, C. and Hand, K. P. and Jia, X.},
  journal = {Planetary Science Journal},
  volume = {3},
  number = {1},
  pages = {5},
  year = {2022},
  publisher = {IOP Publishing}
}

@article{trumbo2019sodium,
  title = {Sodium chloride on the surface of {E}uropa},
  author = {Trumbo, S. K. and Brown, M. E. and Hand, K. P.},
  journal = {Science Advances},
  volume = {5},
  number = {6},
  pages = {eaaw7123},
  year = {2019},
  publisher = {American Association for the Advancement of Science}
}

@article{pappalardo1999does,
  title = {Does {E}uropa Have a Subsurface Ocean? Evaluation of the Geological Evidence},
  author = {Pappalardo, R. T. and Belton, M. J. S. and Breneman, H. H. and Carr, M. H. and Chapman, C. R. and Collins, G. C. and Denk, T. and Fagents, S. and Geissler, P. E. and Giese, B. and others},
  journal = {Journal of Geophysical Research: Planets},
  volume = {104},
  number = {E10},
  pages = {24015--24055},
  year = {1999},
  publisher = {Wiley Online Library}
}

@article{chyba2002europa,
  title = {{E}uropa as an Abode of Life},
  author = {Chyba, C. F. and Phillips, C. B.},
  journal = {Origins of Life and Evolution of the Biosphere},
  volume = {32},
  pages = {47--67},
  year = {2002},
  publisher = {Springer}
}

@article{strazzulla1992ion,
  title = {Ion-Beam--Induced Amorphization of Crystalline Water Ice},
  author = {Strazzulla, G. and Baratta, G. A. and Leto, G. and Foti, G.},
  journal = {Europhysics Letters},
  volume = {18},
  number = {6},
  pages = {517--520},
  year = {1992},
  publisher = {IOP Publishing}
}

@article{trumbo2023distribution,
  title = {The Distribution of {CO$_2$} on {E}uropa Indicates an Internal Source of Carbon},
  author = {Trumbo, S. K. and Brown, M. E.},
  journal = {Science},
  volume = {381},
  number = {6664},
  pages = {1308--1311},
  year = {2023},
  publisher = {American Association for the Advancement of Science}
}

@article{villanueva2023endogenous,
  title = {Endogenous {CO$_2$} Ice Mixture on the Surface of {E}uropa and No Detection of Plume Activity},
  author = {Villanueva, G. L. and Hammel, H. B. and Milam, S. N. and Faggi, S. and Kofman, V. and Roth, L. and Hand, K. P. and Paganini, L. and Stansberry, J. and Spencer, J. and others},
  journal = {Science},
  volume = {381},
  number = {6664},
  pages = {1305--1308},
  year = {2023},
  publisher = {American Association for the Advancement of Science}
}

@article{pavlov2024radiolytic,
  title = {Radiolytic Effects on Biological and Abiotic Amino Acids in Shallow Subsurface Ices on {E}uropa and {E}nceladus},
  author = {Pavlov, A. A. and McLain, H. and Glavin, D. P. and Elsila, J. E. and Dworkin, J. and House, C. H. and Zhang, Z.},
  journal = {Astrobiology},
  volume = {24},
  number = {7},
  pages = {698--709},
  year = {2024},
  publisher = {Mary Ann Liebert, Inc.}
}

@article{brown2013salts,
  title = {Salts and radiation products on the surface of {E}uropa},
  author = {Brown, M. and Hand, K.},
  journal = {Astronomical Journal},
  volume = {145},
  number = {4},
  pages = {110},
  year = {2013},
  publisher = {IOP Publishing}
}

@article{mitchell2017porosity,
  title = {Porosity effects on crystallization kinetics of amorphous solid water: Implications for cold icy objects in the outer solar system},
  author = {Mitchell, E. H. and Raut, U. and Teolis, B. D. and Baragiola, R. A.},
  journal = {Icarus},
  volume = {285},
  pages = {291--299},
  year = {2017},
  publisher = {Elsevier}
}

@article{carlson1999sulfuric,
  title = {Sulfuric acid on {E}uropa and the radiolytic sulfur cycle},
  author = {Carlson, R. and Johnson, R. and Anderson, M.},
  journal = {Science},
  volume = {286},
  number = {5437},
  pages = {97--99},
  year = {1999},
  publisher = {American Association for the Advancement of Science}
}

@article{costello2021impact,
  title={Impact gardening on {E}uropa and repercussions for possible biosignatures},
  author = {Costello, E. and Phillips, C. B. and Lucey, P. and Ghent, R.},
  journal=nata,
  volume={5},
  number={9},
  pages={951--956},
  year={2021},
  publisher={Nature Publishing Group UK London}
}

@article{hansen2004amorphous,
  title={Amorphous and crystalline ice on the Galilean satellites: A balance between thermal and radiolytic processes},
  author = {Hansen, G. B. and McCord, T. B.},
  journal=jgrp,
  volume={109},
  number={E1},
  year={2004},
  publisher={Wiley Online Library}
}

@article{thelen2024subsurface,
  title={Subsurface Thermophysical Properties of {E}uropa’s Leading and Trailing Hemispheres as Revealed by {ALMA}},
  author = {Thelen, A. E. and de Kleer, K. and Camarca, M. and Akins, A. and Gurwell, M. and Butler, B. and de Pater, I.},
  journal=plscij,
  volume={5},
  number={2},
  pages={56},
  year={2024},
  publisher={IOP Publishing}
}

@article{spencer2018plume,
  title={Plume origins and plumbing: from ocean to surface},
  author = {Spencer, J. and Nimmo, F. and Ingersoll, A. P. and Hurford, T. and Kite, E. and Rhoden, A. and Schmidt, J. and Howett, C.},
  journal={{E}nceladus and the icy moons of Saturn},
  volume={163},
  year={2018},
  publisher={University of Arizona Press Tucson}
}

@article{ligier2016vlt,
  title={{VLT}/{SINFONI} observations of {E}uropa: New insights into the surface composition},
  author={Ligier, N and Poulet, F and Carter, J and Brunetto, R and Gourgeot, F},
  journal={The Astronomical Journal},
  volume={151},
  number={6},
  pages={163},
  year={2016},
  publisher={IOP Publishing}
}

@article{trumbo2019h2o2,
  title={{H$_2$O$_2$} within chaos terrain on {E}uropa’s leading hemisphere},
  author={Trumbo, Samantha K and Brown, Michael E and Hand, Kevin P},
  journal={The Astronomical Journal},
  volume={158},
  number={3},
  pages={127},
  year={2019},
  publisher={IOP Publishing}
}

@article{wu2024europa,
  title={{E}uropa’s {H$_2$O$_2$}: Temperature Insensitivity and a Correlation with {CO$_2$}},
  author={Wu, Peiyu and Trumbo, Samantha K and Brown, Michael E and de Kleer, Katherine},
  journal={The Planetary Science Journal},
  volume={5},
  number={10},
  pages={220},
  year={2024},
  publisher={IOP Publishing}
}

@article{loeffler2006synthesis,
  title={Synthesis of hydrogen peroxide in water ice by ion irradiation},
  author={Loeffler, MJ and Raut, U and Vidal, Ricardo Alberto and Baragiola, RA and Carlson, RW},
  journal={Icarus},
  volume={180},
  number={1},
  pages={265--273},
  year={2006},
  publisher={Elsevier}
}

@article{cartwright2025jwst,
  title={{JWST} Reveals Spectral Tracers of Recent Surface Modification on {E}uropa},
  author={Cartwright, Richard J and Hibbits, Charles A and Holler, Bryan J and Raut, Ujjwal and Nordheim, Tom A and Neveu, Marc and Protopapa, Silvia and Glein, Christopher R and Leonard, Erin J and Roth, Lorenz and others},
  journal={arXiv preprint arXiv:2504.05283},
  year={2025}
}

@article{molaro2019microstructural,
  title={The microstructural evolution of water ice in the solar system through sintering},
  author={Molaro, Jamie L and Choukroun, Mathieu and Phillips, Cynthia B and Phelps, Eli S and Hodyss, Robert and Mitchell, Karl L and Lora, Juan M and Meirion-Griffith, Gareth},
  journal={Journal of Geophysical Research: Planets},
  volume={124},
  number={2},
  pages={243--277},
  year={2019},
  publisher={Wiley Online Library}
}

@article{yoffe2025fluorescent,
  title={Fluorescent Biomolecules Detectable in Near-Surface Ice on {E}uropa},
  author={Yoffe, Gideon and Duer-Milner, Keren and Nordheim, Tom Andre and Halevy, Itay and Kaspi, Yohai},
  journal={Astrobiology},
  year={2025},
  publisher={Mary Ann Liebert, Inc., publishers 140 Huguenot Street, 3rd Floor New~…}
}

@article{fama2010radiation,
  title={Radiation-induced amorphization of crystalline ice},
  author={Fam{\'a}, M and Loeffler, MJ and Raut, U and Baragiola, RA},
  journal={Icarus},
  volume={207},
  number={1},
  pages={314--319},
  year={2010},
  publisher={Elsevier}
}

@article{grundy1998temperature,
  title={The temperature-dependent near-infrared absorption spectrum of hexagonal {H$_2$O} ice},
  author={Grundy, WM and Schmitt, B},
  journal={Journal of Geophysical Research: Planets},
  volume={103},
  number={E11},
  pages={25809--25822},
  year={1998},
  publisher={Wiley Online Library}
}

@article{stephan2021vis,
  title={{VIS-NIR/SWIR} spectral properties of {H$_2$O} ice depending on particle size and surface temperature},
  author={Stephan, Katrin and Ciarniello, Mauro and Poch, Olivier and Schmitt, Bernard and Haack, David and Raponi, Andrea},
  journal={Minerals},
  volume={11},
  number={12},
  pages={1328},
  year={2021},
  publisher={MDPI}
}

@article{gerakines2000carbonic,
  title={{C}arbonic acid production in {H$_2$O}: {CO$_2$} ices. {UV} photolysis vs. proton bombardment},
  author={Gerakines, PA and Moore, Marla H and Hudson, Reggie L},
  journal={Astronomy and Astrophysics, v. 357, p. 793-800 (2000)},
  volume={357},
  pages={793--800},
  year={2000}
}

@article{he201812co2,
  title={The $^{12}${CO$_2$} and $^{13}${CO$_2$} absorption bands as tracers of the thermal history of interstellar icy grain mantles},
  author={He, Jiao and Emtiaz, SM and Boogert, Adwin and Vidali, Gianfranco},
  journal={The Astrophysical Journal},
  volume={869},
  number={1},
  pages={41},
  year={2018},
  publisher={IOP Publishing}
}

@article{trumbo2018alma,
  title={{ALMA} thermal observations of {E}uropa},
  author={Trumbo, Samantha K and Brown, Michael E and Butler, Bryan J},
  journal={The Astronomical Journal},
  volume={156},
  number={4},
  pages={161},
  year={2018},
  publisher={IOP Publishing}
}

@article{schiltz2024characterization,
  title={Characterization of carbon dioxide on {G}anymede and {E}uropa supported by experiments: Effects of temperature, porosity, and mixing with water},
  author={Schiltz, L and Escribano, B and Caro, GM Mu{\~n}oz and Cazaux, S and del Burgo Olivares, C and Carrascosa, H and Boszhuizen, I and D{\'\i}az, C Gonz{\'a}lez and Chen, Y-J and Giuliano, BM and others},
  journal={Astronomy \& Astrophysics},
  volume={688},
  pages={A155},
  year={2024},
  publisher={EDP Sciences}
}

@article{mergny2025bond,
  title={A {B}ond albedo map of {E}uropa},
  author={Mergny, C and Schmidt, Fr{\'e}d{\'e}ric and Andrieu, Fran{\c{c}}ois and Belgacem, Ines},
  journal={Astronomy \& Astrophysics},
  volume={693},
  pages={L21},
  year={2025},
  publisher={EDP Sciences}
}

@article{loeffler2020possible,
  title={A possible explanation for the presence of crystalline {H$_2$O}-ice on {K}uiper {B}elt {O}bjects},
  author={Loeffler, Mark J and Tribbett, Patrick D and Cooper, John F and Sturner, Steven J},
  journal={Icarus},
  volume={351},
  pages={113943},
  year={2020},
  publisher={Elsevier}
}

@article{carlson2005distribution,
  title={Distribution of hydrate on {E}uropa: Further evidence for sulfuric acid hydrate},
  author={Carlson, RW and Anderson, MS and Mehlman, R and Johnson, RE},
  journal={Icarus},
  volume={177},
  number={2},
  pages={461--471},
  year={2005},
  publisher={Elsevier}
}

@article{mergny2025blinking,
  title={The blinking crystallinity of {E}uropa: A competition between irradiation and thermal alteration},
  author={Mergny, Cyril and Schmidt, Fr{\'e}d{\'e}ric and Keil, Felix},
  journal={Icarus},
  pages={116700},
  year={2025},
  publisher={Elsevier}
}

@article{mendenhall2005algorithm,
  title={An algorithm for computing screened {C}oulomb scattering in {Geant}4},
  author={Mendenhall, Marcus H and Weller, Robert A},
  journal={Nuclear Instruments and Methods in Physics Research Section B: Beam Interactions with Materials and Atoms},
  volume={227},
  number={3},
  pages={420--430},
  year={2005},
  publisher={Elsevier}
}

@article{nikjoo2016radiation,
  title={Radiation track, {DNA} damage and response—a review},
  author={Nikjoo, H and Emfietzoglou, D and Liamsuwan, T and Taleei, R and Liljequist, David and Uehara, S},
  journal={Reports on Progress in Physics},
  volume={79},
  number={11},
  pages={116601},
  year={2016},
  publisher={IOP Publishing}
}

@article{shin2018development,
  title={Development of a new {Geant4-DNA} electron elastic scattering model for liquid-phase water using the {ELSEPA} code},
  author={Shin, W-G and Bordage, M-C and Emfietzoglou, Dimitris and Kyriakou, Ioanna and Sakata, Dosatsu and Min, CH and Lee, Sang Bae and Guatelli, Susanna and Incerti, Sebastien},
  journal={Journal of Applied Physics},
  volume={124},
  number={22},
  year={2018},
  publisher={AIP Publishing}
}

@article{salvat2005elsepa,
  title={{ELSEPA}—{D}irac partial-wave calculation of elastic scattering of electrons and positrons by atoms, positive ions and molecules},
  author={Salvat, Francesc and Jablonski, Aleksander and Powell, Cedric J},
  journal={Computer physics communications},
  volume={165},
  number={2},
  pages={157--190},
  year={2005},
  publisher={Elsevier}
}

@article{signorell2020electron,
  title={Electron scattering in liquid water and amorphous ice: a striking resemblance},
  author={Signorell, Ruth},
  journal={Physical review letters},
  volume={124},
  number={20},
  pages={205501},
  year={2020},
  publisher={APS}
}

@article{michaud2003cross,
  title={Cross sections for low-energy (1--100 e{V}) electron elastic and inelastic scattering in amorphous ice},
  author={Michaud, M and Wen, A and Sanche, L},
  journal={Radiation research},
  volume={159},
  number={1},
  pages={3--22},
  year={2003}
}

@article{ptasinska2022electron,
  title={Electron scattering processes: fundamentals, challenges, advances, and opportunities},
  author={Ptasinska, Sylwia and Varella, Marcio T do N and Khakoo, Murtadha A and Slaughter, Daniel S and Denifl, Stephan},
  journal={The European Physical Journal D},
  volume={76},
  number={10},
  pages={179},
  year={2022},
  publisher={Springer}
}

@article{afanas2019application,
  title={Application of the photometric theory of the radiance field in the problems of electron scattering},
  author={Afanasyev, Viktor P and Budak, Vladimir P and Efremenko, Dmitry S and Kaplya, Pavel S},
  journal={Light and Engineering},
  volume={27},
  number={2},
  pages={88--96},
  year={2019},
  publisher={Znack Publishing House}
}

@article{emfietzoglou2007consistent,
  title={A consistent dielectric response model for water ice over the whole energy--momentum plane},
  author={Emfietzoglou, D and Nikjoo, H and Petsalakis, Ioannis D and Pathak, A},
  journal={Nuclear Instruments and Methods in Physics Research Section B: Beam Interactions with Materials and Atoms},
  volume={256},
  number={1},
  pages={141--147},
  year={2007},
  publisher={Elsevier}
}

@article{emfietzoglou2005complete,
  title={A complete dielectric response model for liquid water: a solution of the {B}ethe ridge problem},
  author={Emfietzoglou, Dimitris and Cucinotta, Francis A and Nikjoo, Hooshang},
  journal={Radiation research},
  volume={164},
  number={2},
  pages={202--211},
  year={2005}
}

@article{emfietzoglou2017monte,
  title={{M}onte {C}arlo electron track structure calculations in liquid water using a new model dielectric response function},
  author={Emfietzoglou, Dimitris and Papamichael, George and Nikjoo, Hooshang},
  journal={Radiation research},
  volume={188},
  number={3},
  pages={355--368},
  year={2017},
  publisher={The Radiation Research Society}
}

@article{hayashi2000complete,
  title={The complete optical spectrum of liquid water measured by inelastic {X}-ray scattering},
  author={Hayashi, Hisashi and Watanabe, Noboru and Udagawa, Yasuo and Kao, C-C},
  journal={Proceedings of the National Academy of Sciences},
  volume={97},
  number={12},
  pages={6264--6266},
  year={2000},
  publisher={National Academy of Sciences}
}

@article{mermin1970lindhard,
  title={{L}indhard dielectric function in the relaxation-time approximation},
  author={Mermin, N David},
  journal={Physical Review B},
  volume={1},
  number={5},
  pages={2362},
  year={1970},
  publisher={APS}
}

@article{kyriakou2015improvements,
  title={Improvements in {Geant4} energy-loss model and the effect on low-energy electron transport in liquid water},
  author={Kyriakou, I and Incerti, S and Francis, Z},
  journal={Medical physics},
  volume={42},
  number={7},
  pages={3870--3876},
  year={2015},
  publisher={Wiley Online Library}
}

@article{allison2016recent,
  title={Recent developments in {G}eant4},
  author={Allison, John and Amako, Katsuya and Apostolakis, John and Arce, Pedro and Asai, Makoto and Aso, Tsukasa and Bagli, Enrico and Bagulya, A and Banerjee, S and Barrand, GJNI and others},
  journal={Nuclear instruments and methods in physics research section A: Accelerators, Spectrometers, Detectors and Associated Equipment},
  volume={835},
  pages={186--225},
  year={2016},
  publisher={Elsevier}
}

@article{allison2006geant4,
  title={{G}eant4 developments and applications},
  author={Allison, John and Amako, Katsuya and Apostolakis, JEA and Araujo, HAAH and Dubois, P Arce and Asai, MAAM and Barrand, GABG and Capra, RACR and Chauvie, SACS and Chytracek, RACR and others},
  journal={IEEE Transactions on Nuclear Science},
  volume={53},
  number={1},
  pages={270--278},
  year={2006},
  publisher={IEEE}
}

@article{incerti2010comparison,
  title={Comparison of {G}eant4 very low energy cross section models with experimental data in water},
  author={Incerti, S and Ivanchenko, A and Karamitros, M and Mantero, A and Moretto, P and Tran, HN and Mascialino, B and Champion, C and Ivanchenko, VN and Bernal, MA and others},
  journal={Medical physics},
  volume={37},
  number={9},
  pages={4692--4708},
  year={2010},
  publisher={Wiley Online Library}
}

@article{bernal2015track,
  title={Track structure modeling in liquid water: A review of the {G}eant4-{DNA} very low energy extension of the Geant4 Monte Carlo simulation toolkit},
  author={Bernal, Mario A and Bordage, Marie Claude and Brown, Jeremy Michael Cooney and Dav{\'\i}dkov{\'a}, Marie and Delage, E and El Bitar, Z and Enger, Shirin A and Francis, Ziad and Guatelli, Susanna and Ivanchenko, Vladimir N and others},
  journal={Physica Medica},
  volume={31},
  number={8},
  pages={861--874},
  year={2015},
  publisher={Elsevier}
}

@article{incerti2018geant4,
  title={{G}eant4-{DNA} example applications for track structure simulations in liquid water: a report from the {G}eant4-{DNA} {P}roject},
  author={Incerti, Sebastien and Kyriakou, Ioanna and Bernal, MA and Bordage, MC and Francis, Z and Guatelli, Susanna and Ivanchenko, V and Karamitros, M and Lampe, N and Lee, Sang Bae and others},
  journal={Medical physics},
  volume={45},
  number={8},
  pages={e722--e739},
  year={2018},
  publisher={Wiley Online Library}
}

@article{tran2024review,
  title={Review of chemical models and applications in {G}eant4-{DNA}: Report from the {ESA} {B}io{R}ad {III} {P}roject},
  author={Tran, Hoang Ngoc and Archer, Jay and Baldacchino, G{\'e}rard and Brown, Jeremy MC and Chappuis, Flore and Cirrone, Giuseppe Antonio Pablo and Desorgher, Laurent and Dominguez, Naoki and Fattori, Serena and Guatelli, Susanna and others},
  journal={Medical Physics},
  volume={51},
  number={9},
  pages={5873--5889},
  year={2024},
  publisher={Wiley Online Library}
}

@article{incerti2010geant4,
  title={The {G}eant4-{DNA} project},
  author={Incerti, S{\'e}bastien and Baldacchino, G{\'e}rard and Bernal, M and Capra, Riccardo and Champion, Christophe and Francis, Ziad and Gueye, Paul and Mantero, Alfonso and Mascialino, Barbara and Moretto, Philippe and others},
  journal={International Journal of Modeling, Simulation, and Scientific Computing},
  volume={1},
  number={02},
  pages={157--178},
  year={2010},
  publisher={World Scientific}
}

@article{boogert2015observations,
  title={Observations of the icy universe},
  author={Boogert, AC Adwin and Gerakines, Perry A and Whittet, Douglas CB},
  journal={Annual Review of Astronomy and Astrophysics},
  volume={53},
  number={1},
  pages={541--581},
  year={2015},
  publisher={Annual Reviews}
}

@article{hollenbach2008water,
  title={Water, {O}$_2$, and ice in molecular clouds},
  author={Hollenbach, David and Kaufman, Michael J and Bergin, Edwin A and Melnick, Gary J},
  journal={The Astrophysical Journal},
  volume={690},
  number={2},
  pages={1497},
  year={2008},
  publisher={IOP Publishing}
}

@article{herbst1995chemistry,
  title={Chemistry in the interstellar medium},
  author={Herbst, Eric},
  journal={Annual Review of Physical Chemistry},
  volume={46},
  number={1},
  pages={27--54},
  year={1995},
  publisher={Annual Reviews 4139 El Camino Way, PO Box 10139, Palo Alto, CA 94303-0139, USA}
}

@article{potapov2021dust,
  title={Dust/ice mixing in cold regions and solid-state water in the diffuse interstellar medium},
  author={Potapov, Alexey and Bouwman, Jeroen and J{\"a}ger, Cornelia and Henning, Thomas},
  journal={Nature Astronomy},
  volume={5},
  number={1},
  pages={78--85},
  year={2021},
  publisher={Nature Publishing Group UK London}
}

@article{ciesla2006evolution,
  title={The evolution of the water distribution in a viscous protoplanetary disk},
  author={Ciesla, Fred J and Cuzzi, Jeffrey N},
  journal={Icarus},
  volume={181},
  number={1},
  pages={178--204},
  year={2006},
  publisher={Elsevier}
}

@article{kyriakou2025extension,
  title={Extension of the Discrete Electron Transport Capabilities of the {Geant4-DNA} Toolkit to {M}e{V} Energies},
  author={Kyriakou, Ioanna and Tran, Hoang N and Desorgher, Laurent and Ivantchenko, Vladimir and Guatelli, Susanna and Santin, Giovanni and Nieminen, Petteri and Incerti, Sebastien and Emfietzoglou, Dimitris},
  journal={Applied Sciences},
  volume={15},
  number={3},
  pages={1183},
  year={2025},
  publisher={MDPI}
}

@article{king2022compositional,
  title={Compositional mapping of {E}uropa using {MCMC} modeling of near-{IR} {VLT}/{SPHERE} and {G}alileo/{NIMS} observations},
  author={King, Oliver and Fletcher, Leigh N and Ligier, Nicolas},
  journal={The Planetary Science Journal},
  volume={3},
  number={3},
  pages={72},
  year={2022},
  publisher={IOP Publishing}
}

@article{emfietzoglou2003inelastic,
  title={Inelastic cross-sections for electron transport in liquid water: a comparison of dielectric models},
  author={Emfietzoglou, D},
  journal={Radiation Physics and Chemistry},
  volume={66},
  number={6},
  pages={373--385},
  year={2003},
  publisher={Elsevier}
}

@article{emfietzoglou2013inelastic,
  title={Inelastic cross sections for low-energy electrons in liquid water: exchange and correlation effects},
  author={Emfietzoglou, Dimitris and Kyriakou, Ioanna and Garcia-Molina, Rafael and Abril, Isabel and Nikjoo, Hooshang},
  journal={Radiation research},
  volume={180},
  number={5},
  pages={499--513},
  year={2013},
  publisher={The Radiation Research Society}
}

@article{bass2003dissociative,
  title={Dissociative electron attachment and charge transfer in condensed matter},
  author={Bass, Andrew D and Sanche, L{\'e}on},
  journal={Radiation Physics and Chemistry},
  volume={68},
  number={1-2},
  pages={3--13},
  year={2003},
  publisher={Elsevier}
}

@article{sanche1995interactions,
  title={Interactions of low-energy electrons with atomic and molecular solids},
  author={Sanche, Leon},
  journal={Scanning Microscopy},
  volume={9},
  number={3},
  pages={1},
  year={1995}
}

@article{daniels1971bestimmung,
  title={Bestimmung der optischen konstanten von eis aus energie-verlustmessungen von schnellen elektronen},
  author={Daniels, J},
  journal={Optics Communications},
  volume={3},
  number={4},
  pages={240--243},
  year={1971},
  publisher={Elsevier}
}

@article{kobayashi1983optical,
  title={Optical spectra and electronic structure of ice},
  author={Kobayashi, Koichi},
  journal={The Journal of Physical Chemistry},
  volume={87},
  number={21},
  pages={4317--4321},
  year={1983},
  publisher={ACS Publications}
}

@book{petrenko1999physics,
  title={Physics of ice},
  author={Petrenko, Victor F and Whitworth, Robert W},
  year={1999},
  publisher={OUP Oxford}
}

@article{amato2026molecular,
  title={Molecular and pore-scale structure evolution in amorphous solid water},
  author={Amato, Zachary and G{\"a}rtner, Sabrina and Ghesqui{\`e}re, Pierre and Headen, Thomas F and Youngs, Tristan GA and Bowron, Daniel T and Cavalcanti, Leide P and Rogers, Sarah E and Pascual, Natalia and Auriacombe, Olivier and others},
  journal={Physical Chemistry Chemical Physics},
  volume={28},
  number={1},
  pages={524--537},
  year={2026},
  publisher={Royal Society of Chemistry}
}

@article{bhattacharya2014excess,
  title={Excess electrons in ice: a density functional theory study},
  author={Bhattacharya, Somesh Kr and Inam, Fakharul and Scandolo, Sandro},
  journal={Physical Chemistry Chemical Physics},
  volume={16},
  number={7},
  pages={3103--3107},
  year={2014},
  publisher={Royal Society of Chemistry}
}

@techreport{urban2006model,
  title={A model for multiple scattering in {G}eant4},
  author={Urb{\'a}n, L{\'a}szl{\'o}},
  year={2006}
}

@article{apostolakis2010validation,
  title={Validation and verification of {G}eant4 standard electromagnetic physics},
  author={Apostolakis, J and Bagulya, A and Elles, S and Ivanchenko, VN and Jacquemier, J and Maire, M and Toshito, T and Urban, L},
  booktitle={Journal of Physics: Conference Series},
  volume={219},
  number={3},
  pages={032044},
  year={2010}
}

@article{Yoffe_2026,
year = {2026},
month = {apr},
publisher = {The American Astronomical Society},
volume = {1001},
number = {1},
pages = {4},
author = {Yoffe, Gideon and Shahaf, Sahar},
title = {Spectral Decomposition Reveals Surface Processes on Europa},
journal = {The Astrophysical Journal},
}

@article{mergny2024lunaicy,
  title={{LunaIcy}: exploring {E}uropa’s icy surface microstructure through multiphysics simulations},
  author={Mergny, Cyril and Schmidt, Fr{\'e}d{\'e}ric},
  journal={The Planetary Science Journal},
  volume={5},
  number={10},
  pages={216},
  year={2024},
  publisher={The American Astronomical Society}
}

@article{vorburger2018europa,
  title={{E}uropa’s ice-related atmosphere: the sputter contribution},
  author={Vorburger, Audrey and Wurz, Peter},
  journal={Icarus},
  volume={311},
  pages={135--145},
  year={2018},
  publisher={Elsevier}
}

@article{snodgrass2017main,
  title={The main belt comets and ice in the solar system},
  author={Snodgrass, Colin and Agarwal, Jessica and Combi, Michael and Fitzsimmons, Alan and Guilbert-Lepoutre, Aurelie and Hsieh, Henry H and Hui, Man-To and Jehin, Emmanuel and Kelley, Michael SP and Knight, Matthew M and others},
  journal={The Astronomy and Astrophysics Review},
  volume={25},
  number={1},
  pages={5},
  year={2017},
  publisher={Springer}
}

@article{brown2007collisional,
  title={A collisional family of icy objects in the {K}uiper belt},
  author={Brown, Michael E and Barkume, Kristina M and Ragozzine, Darin and Schaller, Emily L},
  journal={Nature},
  volume={446},
  number={7133},
  pages={294--296},
  year={2007},
  publisher={Nature Publishing Group UK London}
}

@article{platz2016surface,
  title={Surface water-ice deposits in the northern shadowed regions of {C}eres},
  author={Platz, T and Nathues, A and Schorghofer, N and Preusker, Frank and Mazarico, E and Schr{\"o}der, SE and Byrne, S and Kneissl, T and Schmedemann, N and Combe, J-P and others},
  journal={Nature Astronomy},
  volume={1},
  number={1},
  pages={0007},
  year={2016},
  publisher={Nature Publishing Group UK London}
}

@article{jenniskens1994structural,
  title={Structural transitions in amorphous water ice and astrophysical implications},
  author={Jenniskens, Peter and Blake, David F},
  journal={Science},
  volume={265},
  number={5173},
  pages={753--756},
  year={1994},
  publisher={American Association for the Advancement of Science}
}

@article{sori2019islands,
  title={Islands of ice on {M}ars and {P}luto},
  author={Sori, Michael M and Bapst, Jonathan and Becerra, Patricio and Byrne, Shane},
  journal={Journal of Geophysical Research: Planets},
  volume={124},
  number={10},
  pages={2522--2542},
  year={2019},
  publisher={Wiley Online Library}
}

@article{hayne2021micro,
  title={Micro cold traps on the {M}oon},
  author={Hayne, Paul O and Aharonson, Oded and Sch{\"o}rghofer, Norbert},
  journal={Nature Astronomy},
  volume={5},
  number={2},
  pages={169--175},
  year={2021},
  publisher={Nature Publishing Group UK London}
}

@article{seltzer1985bremsstrahlung,
  title={Bremsstrahlung spectra from electron interactions with screened atomic nuclei and orbital electrons},
  author={Seltzer, Stephen M and Berger, Martin J},
  journal={Nuclear Instruments and Methods in Physics Research Section B: Beam Interactions with Materials and Atoms},
  volume={12},
  number={1},
  pages={95--134},
  year={1985},
  publisher={Elsevier}
}

@article{seltzer1986bremsstrahlung,
  title={Bremsstrahlung energy spectra from electrons with kinetic energy 1 keV--10 GeV incident on screened nuclei and orbital electrons of neutral atoms with Z= 1--100},
  author={Seltzer, Stephen M and Berger, Martin J},
  journal={Atomic data and nuclear data tables},
  volume={35},
  number={3},
  pages={345--418},
  year={1986},
  publisher={Elsevier}
}

@article{navas2024review,
  title={Review of particle physics},
  author={Navas, Sea and Amsler, C and Gutsche, T and Hanhart, C and Louren{\c{c}}o, C and Masoni, A and Mikhasenko, M and Mitchell, R and Patrignani, C and Schwanda, C and others},
  journal={Physical Review D},
  volume={110},
  number={3},
  year={2024}
}

@article{pines1952collective,
  title={A collective description of electron interactions: {II}. Collective vs individual particle aspects of the interactions},
  author={Pines, David and Bohm, David},
  journal={Physical Review},
  volume={85},
  number={2},
  pages={338},
  year={1952},
  publisher={APS}
}

@article{bethe1930theorie,
  title={Zur theorie des durchgangs schneller korpuskularstrahlen durch materie},
  author={Bethe, Hans},
  journal={Annalen der Physik},
  volume={397},
  number={3},
  pages={325--400},
  year={1930},
  publisher={Wiley Online Library}
}

@article{emfietzoglou2002inelastic,
  title={Inelastic collision characteristics of electrons in liquid water},
  author={Emfietzoglou, D and Moscovitch, M},
  journal={Nuclear Instruments and Methods in Physics Research Section B: Beam Interactions with Materials and Atoms},
  volume={193},
  number={1-4},
  pages={71--78},
  year={2002},
  publisher={Elsevier}
}

@article{dingfelder1998electron,
  title={Electron inelastic-scattering cross sections in liquid water},
  author={Dingfelder, Michael and Hantke, Detlev and Inokuti, Mitio and Paretzke, Herwig G},
  journal={Radiation physics and chemistry},
  volume={53},
  number={1},
  pages={1--18},
  year={1998},
  publisher={Elsevier}
}

@article{kyriakou2021review,
  title={Review of the {Geant4-DNA} simulation toolkit for radiobiological applications at the cellular and {DNA} level},
  author={Kyriakou, Ioanna and Sakata, Dousatsu and Tran, Hoang Ngoc and Perrot, Yann and Shin, Wook-Geun and Lampe, Nathanael and Zein, Sara and Bordage, Marie Claude and Guatelli, Susanna and Villagrasa, Carmen and others},
  journal={Cancers},
  volume={14},
  number={1},
  pages={35},
  year={2021},
  publisher={MDPI}
}

@article{shinotsuka2017calculations,
  title={Calculations of electron inelastic mean free paths. {XI}. {D}ata for liquid water for energies from 50~{eV} to 30~{keV}},
  author={Shinotsuka, Hiroshi and Da, Bo and Tanuma, Shigeo and Yoshikawa, Hideki and Powell, Cedric J and Penn, David R},
  journal={Surface and Interface Analysis},
  volume={49},
  number={4},
  pages={238--252},
  year={2017},
  publisher={Wiley Online Library}
}

@software{icymoonsrelease,
  author       = {Yoffe, Gideon and Pienaar, Jacques},
  title        = {GEANT4-IcyMoons},
  year         = {2026},
  publisher    = {Zenodo},
  version      = {v1.0.0},
  doi          = {10.5281/zenodo.19386442},
  url          = {https://doi.org/10.5281/zenodo.19386442}
}

@article{lunine2006origin,
  title={Origin of water ice in the solar system},
  author={Lunine, Jonathan I},
  journal={Meteorites and the early solar system II},
  pages={309--319},
  year={2006},
  publisher={University of Arizona Press Tucson, AZ}
}

@article{lunine2017ocean,
  title={Ocean worlds exploration},
  author={Lunine, Jonathan I},
  journal={Acta Astronautica},
  volume={131},
  pages={123--130},
  year={2017},
  publisher={Elsevier}
}

@article{lunine1982formation,
  title={Formation of the {G}alilean satellites in a gaseous nebula},
  author={Lunine, Jonathan I and Stevenson, David J},
  journal={Icarus},
  volume={52},
  number={1},
  pages={14--39},
  year={1982},
  publisher={Elsevier}
}
\bibliographystyle{aasjournalv7}

\appendix

\section{Vibrational excitation channels parameters} \label{app:vibparam}

\begin{table*}[h!]
\centering
\caption{Vibrational excitation channels of amorphous ice at 14~K, showing the mean energy loss $E^{(\rm v)}_i$ and full width at half maximum (FWHM) $\Delta_i$ for each mode, adopted from \citet{michaud2003cross}.}
\label{tab:vibparams}
\begin{tabular}{@{}llcc@{}}
\toprule
Category & Mode Description & $E^{(\rm v)}_i$ (eV) & $\Delta_i$ (eV) \\
\midrule
\multirow{3}{*}{\textbf{Intermolecular}} 
& Translational phonon $v_{\mathrm{T}}''$ & 0.024 & 0.025 \\
& Libration $v_{\mathrm{L}}'$              & 0.061 & 0.030 \\
& Libration $v_{\mathrm{L}}''$             & 0.092 & 0.040 \\
\midrule
\multirow{5}{*}{\textbf{Intramolecular}} 
& Bending $v_2$                           & 0.204 & 0.016 \\
& Stretching $v_{1,3}$                    & 0.417 & 0.050 \\
& Asymmetric stretch $v_3$                & 0.460 & 0.005 \\
& Stretch--libration combination $v_{1,3}+v_{\mathrm{L}}$ & 0.510 & 0.040 \\
& Overtone $2v_{1,3}$                     & 0.834 & 0.075 \\
\bottomrule
\end{tabular}
\end{table*}

\section{Inverse transform sampling of scattering angles}\label{app:HG_numerical}

To generate physically accurate scattering angles for each vibrational excitation mode, we employ a probabilistic sampling scheme based on the inverse transform method.  
The goal is to sample deflection angles $\theta$ from an angular probability density function (PDF) $f(\theta)$ defined over the solid angle, such that the statistical ensemble of directions reproduces the desired anisotropy of scattering.

\subsection{Angular probability density}

For each vibrational or elastic scattering channel, the angular deflection probability is described by the Henyey-Greenstein (HG) phase function, which provides a continuous, normalized representation of anisotropic scattering.  
The angular probability density over solid angle is
\begin{equation}
f(\theta; g) = \frac{1 - g^2}{4\pi [1 + g^2 - 2g\cos\theta]^{3/2}},
\end{equation}
where the parameter $g = \langle \cos\theta \rangle$ represents the mean cosine of the scattering angle and controls the degree of forward-peaked behavior.  
When $g=0$, the distribution is isotropic ($f = 1/4\pi$), while $g\rightarrow1$ corresponds to strongly forward scattering.  
For each $i^{\rm{th}}$ mode and incident energy $T$, the value of $g_i(T)$ is chosen such that the HG distribution reproduces the forward-to-backward scattering asymmetry measured experimentally.  
\citet{michaud2003cross} reported a hemispheric anisotropy coefficient $\gamma_i(T)$, which defines the relative scattering probability in the forward ($0\le\theta\le\pi/2$) and backward ($\pi/2\le\theta\le\pi$) hemispheres.  
The HG asymmetry parameter is obtained by requiring the integrated forward fraction of the HG kernel to match the measured value:
\begin{equation}
\int_0^{\pi/2} f(\theta; g_i)\,2\pi\sin\theta\,d\theta
=\frac{1+\gamma_i(T)}{2}.
\end{equation}
This condition defines a monotonic mapping $\gamma_i \mapsto g_i$, solved numerically for each mode and energy. It spans the full range from isotropic scattering ($g_i=0$) to the forward-scattering limit ($g_i \to 1$). The resulting $g_i(T)$ values match the measured hemispheric anisotropy and enable smooth, energy-dependent HG angular sampling for each process.

\subsection{Cumulative distribution function}

The cumulative distribution function (CDF) corresponding to the HG kernel is defined as
\begin{equation}
F(\mu; g) = \int_{-1}^{\mu} f(\theta; g)\, 2\pi \sin\theta\, d\theta,
\qquad \text{with } \mu = \cos\theta.
\end{equation}
After integration, the analytic form of $F(\mu; g)$ is
\begin{equation}
F(\mu; g) = 
\frac{1 - g^2}{2g}
\left[
\frac{1}{1 - g}
- \frac{1}{\sqrt{1 + g^2 - 2g\mu}}
\right],
\end{equation}
which monotonically increases from 0 at $\mu=-1$ to 1 at $\mu=+1$.  
This provides a direct mapping between a random number $\xi \in [0,1]$ and the cosine of the scattering angle $\mu$.

\subsection{Inverse transform sampling}

The inverse transform method proceeds as follows:

\begin{enumerate}
    \item Draw a uniform random number $\xi \in [0,1]$.
    \item Set $F(\mu; g) = \xi$ and solve for $\mu = \cos\theta$.
\end{enumerate}

For the HG function, the inversion can be expressed analytically:
\begin{equation}
\mu = 
\frac{1}{2g}
\left[
1 + g^2 -
\left(
\frac{1 - g^2}{1 - g + 2g\xi}
\right)^2
\right].
\end{equation}
The deflection angle is then obtained as $\theta = \arccos(\mu)$.  
This provides an exact and efficient sampling rule for the HG distribution.

In practice, the above expression is used to generate scattering angles during Monte Carlo transport simulations.  
For verification and compatibility with tabulated data, the cumulative distributions $F(\theta; g_i(T))$ are also computed numerically on a dense grid of $\theta$ values using the integral form of Equation~(A3).  
At runtime, when $\xi$ is drawn, the inverse transform is evaluated by interpolating within this precomputed monotonic table, ensuring stable and accurate sampling across all scattering regimes.

\section{Electronic excitation and ionization} \label{app:excit_ion}

\subsection{Drude kernels and finite momentum coefficients}\label{app:drude}

The individual Drude kernels are defined as
\begin{subequations}
\begin{align}
D_2(E; f_i, E_{0,i}, \Gamma_i)
&=\frac{f_i\,\Gamma_i E}
{(E_{0,i}^2 - E^2)^2 + (\Gamma_i E)^2},
\label{eq:D2a}\\[4pt]
D'_2(E; f_j, E_{0,j}, \Gamma_j)
&=\frac{2 f_j\,\Gamma_j^3 E^3}
{\bigl[(E_{0,j}^2 - E^2)^2 + (\Gamma_j E)^2\bigr]^2},
\label{eq:D2b}\\[4pt]
D_1(E; f_i, E_{0,i}, \Gamma_i)
&=\frac{f_i\,(E_{0,i}^2 - E^2)}
{(E_{0,i}^2 - E^2)^2 + (\Gamma_i E)^2},
\label{eq:D1a}\\[4pt]
D'_1(E; f_j, E_{0,j}, \Gamma_j)
&=\frac{f_j\,(E_{0,j}^2 - E^2)}
{\bigl[(E_{0,j}^2 - E^2)^2 + (\Gamma_j E)^2\bigr]^2}\notag\\
&\quad\times\Bigl[(E_{0,j}^2 - E^2)^2 + 3(\Gamma_j E)^2\Bigr].
\label{eq:D1b}
\end{align}
\end{subequations}

These standard and derivative Drude terms yield a self-consistent and causal dielectric response, where $\varepsilon_1$ and $\varepsilon_2$ satisfy the $f$-sum rules for $\mathrm{Im}\,\varepsilon$ and $\mathrm{Im}(1/\varepsilon)$ to within 1\%, as verified for both amorphous and hexagonal ice \citep{emfietzoglou2007consistent}.  
The fitted oscillator parameters for all excitation and ionization channels used in this model are listed in Table~\ref{tab:drudeparams_thresholds_combined}.

\begin{table*}[h!]
\centering
\caption{Optical-limit Drude--oscillator parameters and threshold values used for amorphous and hexagonal ice. For each channel, we list the central energy $E_0$, damping width $\Gamma$, oscillator strength $f$, and threshold $B_{\mathrm{th}}$, adopted for water-ice values adopted from \citet{emfietzoglou2007consistent}.}
\label{tab:drudeparams_thresholds_combined}
\begin{tabular}{@{}llcccccccc@{}}
\toprule
\multirow{2}{*}{\textbf{Channel type}} &
\multirow{2}{*}{\textbf{Channel description}} &
\multicolumn{4}{c}{\textbf{Amorphous ice}} &
\multicolumn{4}{c}{\textbf{Hexagonal ice}} \\
\cmidrule(lr){3-6} \cmidrule(l){7-10}
&
& $E_0$ (eV) & $\Gamma$ (eV) & $f$ & $B_{\mathrm{th}}$ (eV)
& $E_0$ (eV) & $\Gamma$ (eV) & $f$ & $B_{\mathrm{th}}$ (eV) \\
\midrule
Excitation onset & $B_{\min}$ 
& --- & --- & --- & 7.0 
& --- & --- & --- & 7.0 \\
\midrule
Excitation & $A_1 \!\leftrightarrow\! B_1$ 
& 8.65 & 1.6 & 0.0090 & --- 
& 8.65 & 1.6 & 0.0168 & --- \\
Excitation & $B_1 \!\leftrightarrow\! A_1$ 
& 10.50 & 2.5 & 0.0096 & --- 
& 10.50 & 1.5 & 0.0065 & --- \\
Excitation & Rydberg A+B                    
& 12.60 & 3.5 & 0.0210 & --- 
& 12.60 & 3.0 & 0.0190 & --- \\
Excitation & Rydberg C+D                    
& 14.10 & 3.0 & 0.0040 & --- 
& 14.10 & 2.7 & 0.0110 & --- \\
Excitation & Diffuse band                   
& 14.50 & 2.5 & 0.0030 & --- 
& 14.50 & 1.5 & 0.0044 & --- \\
\midrule
Ionization & $1b_1$                         
& 15.40 & 5.7  & 0.1250 & 10.0 
& 15.80 & 4.6  & 0.1000 & 10.0 \\
Ionization & $3a_1$                         
& 18.60 & 7.1  & 0.1300 & 13.0 
& 18.00 & 7.5  & 0.2000 & 13.0 \\
Ionization & $1b_2$                         
& 24.50 & 15.0 & 0.1100 & 17.0 
& 24.50 & 14.0 & 0.1100 & 17.0 \\
Ionization & $2a_1$                         
& 38.00 & 30.0 & 0.4110 & 32.0 
& 35.00 & 30.0 & 0.3580 & 32.0 \\
\midrule
K-shell & O K-shell                         
& 450.0 & 360.0 & 0.3143 & 540.0 
& 450.0 & 360.0 & 0.3143 & 540.0 \\
\bottomrule
\end{tabular}
\end{table*}

The per-excitation dispersion triplets $(a_j,b_j,c_j)$ and the global coefficients $(c_{\mathrm{disp}},d_{\mathrm{disp}},b_1,b_2)$ are listed in Table~\ref{tab:finiteq-coeffs}; they follow the ECN calibration and are applied identically to amorphous and hexagonal ice in the implementation within Geant4-IcyMoons \citep{emfietzoglou2017monte}.

\begin{table*}[h!]
\centering
\caption{Finite-$q$ dispersion coefficients for the ECN scheme. The form of the ice finite-$q$ extension follows \citet{emfietzoglou2007consistent}, while the global ionization-energy and lifetime-broadening coefficients $(c_{\mathrm{disp}},d_{\mathrm{disp}},b_1,b_2)$ follow the ECN liquid-water calibration of \citet{emfietzoglou2017monte}. The per-excitation triplets $(a_j,b_j,c_j)$ are adopted from \citet{dingfelder1998electron}.}
\label{tab:finiteq-coeffs}
\begin{tabular}{@{}llccc@{}}
\toprule
Category & Band / Parameter & $a_j$ & $b_j$ & $c_j$ \\
\midrule
\multirow{5}{*}{\textbf{Excitations}}
& $j=1$ (A$_1\!\leftrightarrow\!$ B$_1$) & 3.82 & 0.0272 & 0.098 \\
& $j=2$ (B$_1\!\leftrightarrow\!$ A$_1$) & 2.47 & 0.0295 & 0.075 \\
& $j=3$ (Ryd.~A+B)                      & 2.47 & 0.0311 & 0.074 \\
& $j=4$ (Ryd.~C+D)                      & 3.01 & 0.0111 & 0.765 \\
& $j=5$ (Diffuse band)                  & 2.44 & 0.0633 & 0.425 \\
\midrule
\textbf{Global} & $c_{\mathrm{disp}}$ (energy shift) & \multicolumn{3}{c}{1.5} \\
\textbf{coeffs} & $d_{\mathrm{disp}}$ (energy shift) & \multicolumn{3}{c}{0.4} \\
                & $b_{1}$ (linear broadening)        & \multicolumn{3}{c}{0.735} \\
                & $b_{2}$ (quadratic broadening)     & \multicolumn{3}{c}{0.441} \\
\bottomrule
\end{tabular}
\end{table*}

\subsection{Excitation--ionization partitioning correction}\label{app:partitioning_fix}

To maintain a consistent separation between excitation and ionization channels in the optical dielectric function, we adopt the partitioning formalism of Eqs.~(A1)--(A9) of \citet{kyriakou2015improvements}, while distinguishing explicitly between the minimum binding energy $B_1$ and the valence onset $B_{\min}$. The purpose of the scheme is to truncate ionization tails at finite energy, reassign the removed strength to permitted excitation channels, and conserve the total imaginary part of the dielectric function at every energy.
In the optical limit, the imaginary part of the dielectric function is written as
\begin{equation}
\varepsilon_2(E; q{=}0)
=
\sum_{j} \varepsilon_{2,j}(E)
+
\sum_{i} \varepsilon_{2,i}(E),
\end{equation}
where $j$ indexes excitation channels and $i$ indexes ionization shells. Each ionization shell has a threshold $B_i$, with $B_1 \equiv \min_i B_i$ the minimum binding energy, while the excitation manifold begins at the valence onset $B_{\min}$.

Following \citet{kyriakou2015improvements}, the ionization components are modified according to
\begin{equation}
\begin{aligned}
\varepsilon_{2,i}^{\mathrm{(mod)}}(E)
&=
\big[\varepsilon_{2,i}(E)+S_i(E)+C_i^{>}(E)\big]\\
&\quad\times H(E-B_i)\,.
\end{aligned}
\end{equation}

where $H$ is the Heaviside function. The redistribution terms $S_i(E)$ and $C_i^{>}(E)$ reallocate truncated ionization strength among the ionization shells so that $\sum_i \varepsilon_{2,i}^{\mathrm{(mod)}}(E)$ is conserved, as in Eqs.~(A1)--(A5) of \citet{kyriakou2015improvements}.

The excitation channels are modified as
\begin{equation}\label{eq:exc_mod_j}
\begin{aligned}
\varepsilon_{2,j}^{\mathrm{(mod)}}(E)
&=\big[\varepsilon_{2,j}(E)+S_{1j}(E)+C_j(E)\big]\\
&\quad\times H(E-B_{\min})\,,
\end{aligned}
\end{equation}
where $H(E-B_{\min})$ applies the excitation gate at the valence onset.

The first excitation redistribution term transfers the high-energy tail of the lowest ionization shell ($i{=}1$) into the excitation manifold:
\begin{equation}\label{eq:Sn1_j}
\begin{aligned}
S_{1j}(E)
&=\varepsilon_{2,1}(B_1)\,
\exp(B_1 - E)\,
H(E - B_1)\\
&\quad\times
\frac{\varepsilon_{2,j}(E)}
{\sum_{m} \varepsilon_{2,m}(E)}\,.
\end{aligned}
\end{equation}
Here $\varepsilon_{2,1}(B_1)$ is the value of the first ionization shell at threshold, and the ratio distributes this strength among excitation channels in proportion to their instantaneous optical weights at energy $E$.

The second excitation redistribution term assigns ionization strength lying below $B_1$ to excitation channels whose characteristic energies satisfy $E_{0,j}\le B_1$:
\begin{equation}\label{eq:Cj}
\begin{aligned}
C_j(E)
&=\Bigg[\sum_{i}\varepsilon_{2,i}(E)\,H(B_1-E)\Bigg]\\
&\quad\times
\frac{\varepsilon_{2,j}(E)\,H(B_1-E_{0,j})}
{\sum_{m}\varepsilon_{2,m}(E)\,H(B_1-E_{0,m})}\,.
\end{aligned}
\end{equation}
The bracketed factor is the ionization strength available for reassignment at energies $E<B_1$, and the ratio restricts redistribution to the subset of excitation channels allowed by $E_{0,j}\le B_1$.

By construction, $B_1$ appears only in the redistribution terms $S_{1j}$ and $C_j$, whereas $B_{\min}$ appears only in the excitation gate of Equation~\eqref{eq:exc_mod_j}. The total imaginary part after partitioning is
\begin{equation}
\varepsilon_2^{\mathrm{(mod)}}(E)
=
\sum_{j} \varepsilon_{2,j}^{\mathrm{(mod)}}(E)
+
\sum_{i} \varepsilon_{2,i}^{\mathrm{(mod)}}(E),
\end{equation}
which equals the original $\varepsilon_2(E)$ at all energies.

Beyond the optical limit, the same partitioning is applied at finite momentum transfer $q$. The excitation and ionization components, $\varepsilon_{2,j}(E,q)$ and $\varepsilon_{2,i}(E,q)$, are first evaluated using the momentum-dependent Drude parameters $E_n(q)$, $\Gamma_n(q)$, and $f_n(q)$ defined by the dispersion relations of \citet{emfietzoglou2007consistent}. While the spectral peaks shift and broaden with increasing $q$, the physical thresholds $B_i$ and the valence onset $B_{\min}$ remain fixed. The redistribution terms $S_i(E,q)$, $C_i^{>}(E,q)$, $S_{1j}(E,q)$, and $C_j(E,q)$ are recomputed at each $q$ using the dispersed spectral shapes, enforcing vanishing energy loss below threshold throughout the $(E,q)$ plane and accounting for the channel-dependent $q$ evolution of the response \citep{kyriakou2015improvements}.

\subsection{Relativistic corrections for excitation and ionization differential cross-sections}
\label{app:rel_corrections}

To extend the dielectric cross-sections beyond the non-relativistic domain while keeping the
implementation economical, we follow the strategy of applying each correction only where its
impact on the electronic stopping power exceeds the $1\%$ level 
\citep[See Figure 1 of][]{kyriakou2025extension}. The model is partitioned into four incident-energy
regimes, with transitions chosen according to the stopping-power criterion. Using the definitions adopted in \citet{kyriakou2025extension}, regime~II is $1$--$100~\mathrm{keV}$,
regime~III is $0.1$--$1~\mathrm{MeV}$, and regime~IV is $1$--$10~\mathrm{MeV}$. regime~I
corresponds to the remaining low-energy interval down to the model threshold at
$10~\mathrm{eV}$, below $1~\mathrm{keV}$. 

\paragraph{Low-energy Mott--Coulomb corrections (Regimes I and II).}
At sub-keV and keV energies, the plane-wave Born approximation (PWBA) requires Mott--Coulomb
type corrections. For ionization of the $i^{\rm th}$ ionization shell, the corrected differential
cross-section is written as 

\begin{equation}\label{eq:mottco}
\begin{multlined}[t]
\frac{d\sigma^{(i)}_{\mathrm{Mott\!-\!Co}}(E;T)}{dE}
=
\frac{d\sigma^{(i)}_{\mathrm{PWBA}}(E;\,T+B_i+U_i)}{dE}
\\
+
\frac{d\sigma^{(i)}_{\mathrm{PWBA}}(T+2B_i+U_i-E;\,T+B_i+U_i)}{dE}
\\
-
\sqrt{
\frac{d\sigma^{(i)}_{\mathrm{PWBA}}(E;\,T+B_i+U_i)}{dE}
}
\\
\times
\sqrt{
\frac{d\sigma^{(i)}_{\mathrm{PWBA}}(T+2B_i+U_i-E;\,T+B_i+U_i)}{dE}
}\,.
\end{multlined}
\end{equation}
Following the adopted regime criterion, these low-energy corrections are applied only in
regimes~I and II. 

\paragraph{Relativistic plane-wave Born approximation (Regimes II to IV).}
In the relativistic plane-wave Born approximation (RPWBA), the inelastic energy-transfer
differential cross-section is written as the sum of a longitudinal and a transverse contribution,
\begin{equation}
\frac{d\sigma_{\mathrm{RPWBA}}(E;T)}{dE}
=
\frac{d\sigma_{\mathrm{L}}(E;T)}{dE}
+
\frac{d\sigma_{\mathrm{T}}(E;T)}{dE}.
\label{eq:rpwba_split}
\end{equation}
The dielectric response enters through the energy-loss function (ELF),
\begin{equation}
\mathrm{ELF}(E,q)\equiv \Im\!\left[-\frac{1}{\varepsilon(E,q)}\right].
\label{eq:elf_def}
\end{equation}

\paragraph{Longitudinal RPWBA term.}
The longitudinal term is obtained from a momentum-transfer integration of the ELF, but differs
from the non-relativistic PWBA expression due to relativistic kinematics:
\begin{equation}\label{eq:rpwba_long}
\begin{aligned}
\frac{d\sigma_{\mathrm{L}}(E;T)}{dE}
&=
\frac{1}{\pi a_0 N m c^2\,\beta^2(T)}
\\
&\quad\times
\int_{q_{\min,\mathrm{rel}}}^{q_{\max,\mathrm{rel}}}
\frac{c^2 q}{\sqrt{c^2 q^2+(m c^2)^2}}
\,\frac{dq}{Q(q)}
\\
&\quad\times
\frac{1+Q(q)/(m c^2)}{1+Q(q)/(2 m c^2)}
\\
&\quad\times
\mathrm{ELF}(E,q)\,.
\end{aligned}
\end{equation}

Here, $N$ is the molecular number density of the medium and
\begin{equation}
\beta^2(T) \equiv \frac{v^2}{c^2}
=
1-\frac{1}{\left(T/(m c^2)+1\right)^2},
\label{eq:beta_rel}
\end{equation}
while the relativistic free-recoil energy is
\begin{equation}
Q(q)=\sqrt{c^2 q^2+(m c^2)^2}-m c^2.
\label{eq:Q_rel}
\end{equation}
For a given energy transfer $E$, the relativistic kinematic limits of the momentum transfer are
\begin{equation}\label{eq:q_limits_rel}
\begin{aligned}
q_{\max,\mathrm{rel}/\min,\mathrm{rel}}
&=\frac{1}{c}\Bigg[
\sqrt{T(T+2 m c^2)}
\\
&\qquad\pm
\sqrt{(T-E)(T-E+2 m c^2)}
\Bigg].
\end{aligned}
\end{equation}

With the adopted regime logic, the longitudinal RPWBA expression replaces the non-relativistic
PWBA form in regimes~II, III, and IV. In regime~II, the Mott--Coulomb corrections above are
applied to the longitudinal term.

\paragraph{Transverse RPWBA term (Fano small-angle approximation).}
For the transverse term, the small-angle approximation of Fano is used, which makes the result
depend only on the optical limit of the ELF:
\begin{equation}
\begin{aligned}
\frac{d\sigma_{\mathrm{T}}(E;T)}{dE}
&=
\frac{1}{\pi a_0 N m c^2\,\beta^2(T)}
\,
\mathrm{ELF}(E,q=0)
\\
&\qquad\times
\left[
\ln\!\left(\frac{1}{1-\beta^2(T)}\right)-\beta^2(T)
\right].
\end{aligned}
\label{eq:rpwba_trans}
\end{equation}
This transverse contribution vanishes in the non-relativistic limit and is included only in regimes~III and IV under the adopted criterion. 

\paragraph{Fermi density-effect correction (Regime IV only).}
At condensed media and incident energies above the projectile rest-mass energy, distant transverse
interactions are reduced by the density effect. Following \citet{kyriakou2025extension}, we write the density-effect contribution to the transverse inelastic DCS as
\begin{equation}
\frac{d\sigma_{\mathrm{T},\delta}(E;T)}{dE}
=
\frac{1}{\pi a_0 N m c^2\,\beta^2(T)}
\,
\mathrm{ELF}(E,q=0)\,\delta_F(T)\,.
\label{eq:fermi_density_effect}
\end{equation}
This term is considered only in Regime IV, which reduces the transverse inelastic DCS through the enhanced screening of distant
interactions. Thus, the full transverse contribution is written as
\begin{equation}
\frac{d\sigma_{\mathrm{T,IV}}(E;T)}{dE}
=
\frac{d\sigma_{\mathrm{T}}(E;T)}{dE}
-
\frac{d\sigma_{\mathrm{T},\delta}(E;T)}{dE}.
\label{eq:full_transverse_term_regime4}
\end{equation}
Equivalently, the regime~IV transverse contribution may be written in compact form as
\begin{equation}
\frac{d\sigma_{\mathrm{T,IV}}(E;T)}{dE}
=
\frac{1}{\pi a_0 N m c^2\,\beta^2(T)}
\,
\mathrm{ELF}(E,q=0)
\left[
\ln\!\left(\frac{1}{1-\beta^2(T)}\right)-\beta^2(T)-\delta_F(T)
\right].
\label{eq:full_transverse_term_regime4_compact}
\end{equation}
The density-effect parameter $\delta_F$ is determined through the auxiliary function $\delta(X)$,
\begin{align}
\delta_F(T)\equiv \delta(X)
&=
\begin{cases}
4.6052\,X
+\alpha\,(X_1-X)^m
+C, & X_0<X<X_1,\\[4pt]
4.6052\,X
+C, & X>X_1,
\end{cases}
\label{eq:sternheimer_delta}\\[4pt]
X
&=
\log_{10}\!\left(\dfrac{\beta(T)}{\sqrt{1-\beta^2(T)}}\right).
\label{eq:sternheimer_X}
\end{align}

For liquid water, the parameter values used in \citet{kyriakou2025extension} are
$0.09116$, $0.24$, $2.8004$, $3.4773$, and $-3.5017$ for $\alpha,X_0,X_1,m,C$, respectively.
Table~\ref{tab:corrections_regimes} summarizes which terms are active in each incident-energy regime.

\begin{table*}
\tiny
\centering
\caption{Energy-regime logic for the inelastic model corrections.}
\label{tab:corrections_regimes}
\begin{tabular}{llll}
\hline
Regime & $T$ range & Longitudinal term & Additional terms enabled \\
\hline
I   & $10~\mathrm{eV}\le T < 1~\mathrm{keV}$  &
PWBA, Equation~(2) in \citep{kyriakou2025extension} &
Mott--Coulomb corrections, Equation~\eqref{eq:mottco} \\
II  & $1~\mathrm{keV}\le T \le 100~\mathrm{keV}$ &
RPWBA longitudinal, Equation~\eqref{eq:rpwba_long} &
Mott--Coulomb applied to longitudinal term \\
III & $0.1~\mathrm{MeV}\le T \le 1~\mathrm{MeV}$ &
RPWBA longitudinal, Equation~\eqref{eq:rpwba_long} &
Transverse term, Equation~\eqref{eq:rpwba_trans} \\
IV  & $1~\mathrm{MeV}\le T \le 10~\mathrm{MeV}$ &
RPWBA longitudinal, Equation~\eqref{eq:rpwba_long} &
Transverse term plus density effect, Eqs.~\eqref{eq:rpwba_trans}--\eqref{eq:full_transverse_term_regime4_compact} \\
\hline
\end{tabular}
\end{table*}

\subsection{High-energy asymptotic correction to the ELF}

The dielectric model adopted here reproduces the optical and finite-$q$ response of water ice well over the main range relevant for transport, but at large transferred energies, its Drude tail remains too strong relative to the expected asymptotic behavior \citep{kyriakou2025extension}. As a result, the corresponding energy-loss function can overestimate the contribution from large transferred energies. Following the high-energy extension introduced for Geant4-DNA, we apply a mild asymptotic correction to the ELF above a prescribed transferred-energy threshold. This correction is introduced multiplicatively, so that the baseline dielectric construction is preserved at low and moderate transferred energies, while the tail is gradually steepened in the relativistic regime.

Specifically, we define
\begin{equation}
f_{\mathrm{roll}}(E)=
\begin{cases}
1, & E \le E_0, \\[6pt]
1-\alpha \log_{10}\!\left(\dfrac{E}{E_0}\right), & E>E_0,
\end{cases}
\end{equation}
where $E_0 = 50$~keV is the onset energy of the asymptotic correction, and $\alpha=0.05$ sets its strength. We adopt these values following the corresponding treatment for liquid water in \citet{kyriakou2025extension}. The corrected channel-resolved loss function is then
\begin{equation}
\mathrm{ELF}^{(k)}_{\mathrm{corr}}(E,q)
=
f_{\mathrm{roll}}(E)\,\mathrm{ELF}^{(k)}(E,q),
\end{equation}
with the same factor applied to the optical-limit transverse contribution when that term is active.
This correction is asymptotic only. It is not part of the optical fit, but a high-energy modification that suppresses the excess dielectric-loss strength of the Drude tail at large transferred energies. In the present implementation, it is applied only in the high-energy regimes, together with the other relativistic corrections \citep{kyriakou2025extension}.

\section{Simulating electron bombardment of Europa}\label{app:Europa}

\subsection{Numerical implementation}
\label{app:europa_bombardment_numerical}

We discretize Europa’s surface into a $30\times30$ latitude--longitude grid. For each surface cell $k$, the precipitation model assigns an admissible kinetic-energy interval $[T_{\min}^{(k)},T_{\max}^{(k)}]$, specifying which part of the incident electron spectrum can access that location. The local energy flux is therefore
\begin{equation}
\Phi_T^{(k)} =
\int_{T_{\min}^{(k)}}^{T_{\max}^{(k)}} j(T)\,T\,dT.
\end{equation}
Because the transport calculation is performed with monoenergetic primaries, we approximate this continuous integral by a discrete sum over logarithmically sampled energies $\{T_i\}$,
\begin{equation}
\Phi_T^{(k)} \approx
\sum_i j(T_i)\,T_i\,\Delta T_i^{(k)},
\end{equation}
where $\Delta T_i^{(k)}$ is the overlap of the $i^{\rm th}$ with the admissible interval of cell $k$. The number of sampled energies is chosen such that, for every surface cell, this discrete representation reproduces the continuous energy flux to within $1\%$.

Each sampled energy is simulated in a laterally uniform slab of water ice. This idealized geometry isolates the transport physics within the ice, while Europa’s geographic variability is introduced only afterward, when the monoenergetic results are combined with the local spectral weights of each surface cell. In this way, the Geant4 runs provide a common physical library that can be mapped onto the spatially varying bombardment environment across Europa.

Monoenergetic primaries are injected from a point source placed infinitesimally above the ice surface, with the source axis aligned with the local surface normal. The ice is represented as a homogeneous slab centered at $x=y=0$, with half-widths of 2~m in both horizontal directions and thickness 5~m along the depth axis, such that the target occupies $z\in[0,5]$~m. Primaries are initialized at $ z =- 10^ {-3}$ mm and propagate into the ice along the $+z$ direction.

Incident directions are sampled over the downward hemisphere above the local surface. We draw the azimuth uniformly,
\begin{equation}
\phi \sim U(0,2\pi),
\end{equation}
so that no horizontal direction is preferred, and draw the polar angle $\theta$, measured from the surface normal, from a cosine-weighted distribution over $0 \leq \theta \leq \pi/2$. This favors near-normal incidence over grazing trajectories, as expected for particles crossing a planar surface. The injected momentum vector is therefore
\begin{equation}
\mathbf{p} =
\left(
p\sin\theta\cos\phi,\,
p\sin\theta\sin\phi,\,
p\cos\theta
\right).
\end{equation}
Equivalently, the angular distribution may be written as
\begin{equation}
p(\theta,\phi)\,d\theta\,d\phi
=
\frac{\cos\theta\,\sin\theta}{\pi}\,d\theta\,d\phi,
\qquad
0\le\theta\le\frac{\pi}{2},\;\;
0\le\phi<2\pi,
\end{equation}
or, with respect to solid angle,
\begin{equation}
p(\Omega)=\frac{\cos\theta}{\pi}.
\end{equation}
The corresponding hemispheric normalization is
\begin{equation}
\int_{\rm hem.}\cos\theta\,d\Omega
=
\int_{0}^{2\pi}\int_{0}^{\pi/2}\cos\theta\sin\theta\,d\theta\,d\phi
=
\pi.
\end{equation}
Accordingly, if $j(T)$ is the differential intensity in units of
$\#\,\mathrm{cm^{-2}\,s^{-1}\,sr^{-1}\,MeV^{-1}}$,
then the downward flux through the surface is
\begin{equation}
F(T)=\int_{\rm hem.} j(T)\cos\theta\,d\Omega = \pi j(T),
\end{equation}
for an angularly isotropic incident intensity. The $\pi$ factor is therefore purely geometric: it is the conversion between an isotropic intensity per steradian and the net flux through a plane.

For each monoenergetic run, we record the energy deposited as a function of depth, together with the energy leaving the slab in escaping electrons and emitted photons. This yields a vertical irradiation profile for each incident energy. Low-energy electrons deposit most of their energy close to the surface, whereas higher-energy electrons penetrate farther and distribute their energy over greater depths. The full bombardment field across Europa is then reconstructed by combining these monoenergetic depth profiles with the location-dependent spectral weights of each surface cell.

\subsection{Energy loss}
\label{app:energy_loss}

The two four-panel plots decompose the incident electron energy into four recorded channels: energy deposited within the ice, energy returned through the entrance surface by backscattered electrons, energy escaping through the lateral boundaries, and energy leaving through the lower boundary. This partition provides a compact summary of how the incident energy is redistributed during transport. At low incident energies, most of the energy is deposited locally, whereas at higher energies a larger fraction is diverted into escape channels.

Because the simulated ice slab is $5\,\mathrm{m}$ thick, while the deepest electron energy-deposition event in our runs occurs at only $\sim 1.1\,\mathrm{m}$, electrons do not reach the lower boundary. The forward-escaping component is therefore entirely radiative and consists of bremsstrahlung photons produced during transport. The four-channel decomposition thus separates the fraction of the incident energy retained locally from the fraction removed by backscattering, lateral leakage, and radiative escape.



\begin{figure*}
\centering
\rotatebox[origin=c]{0}{\includegraphics[scale = 0.32]{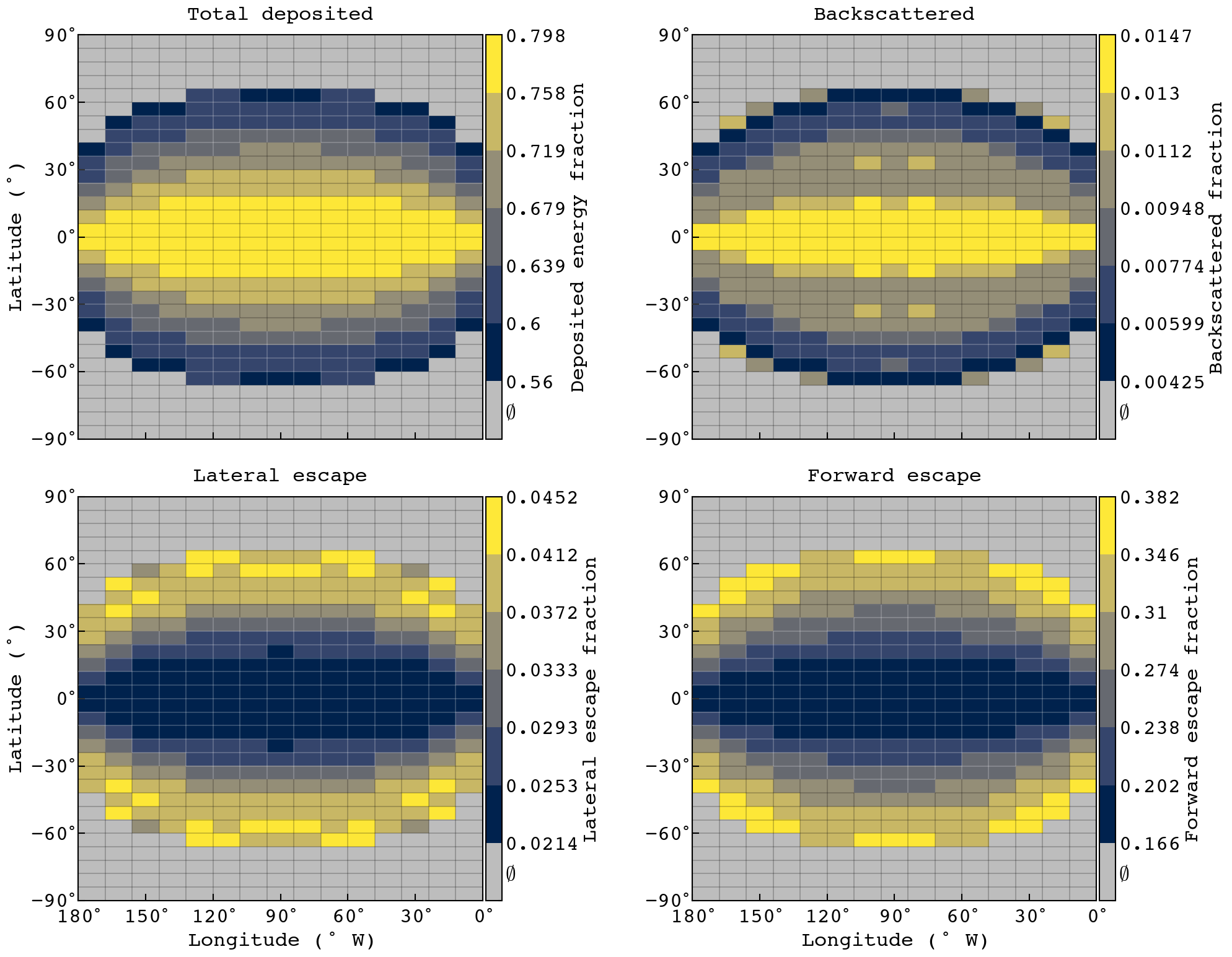}}
  \caption{Energy deposition diagnostics for electron bombardment simulations of Europa's leading hemisphere.}
     \label{fig:europa_4p_trailing_edep}
\end{figure*}

\begin{figure*}
\centering
\rotatebox[origin=c]{0}{\includegraphics[scale = 0.32]{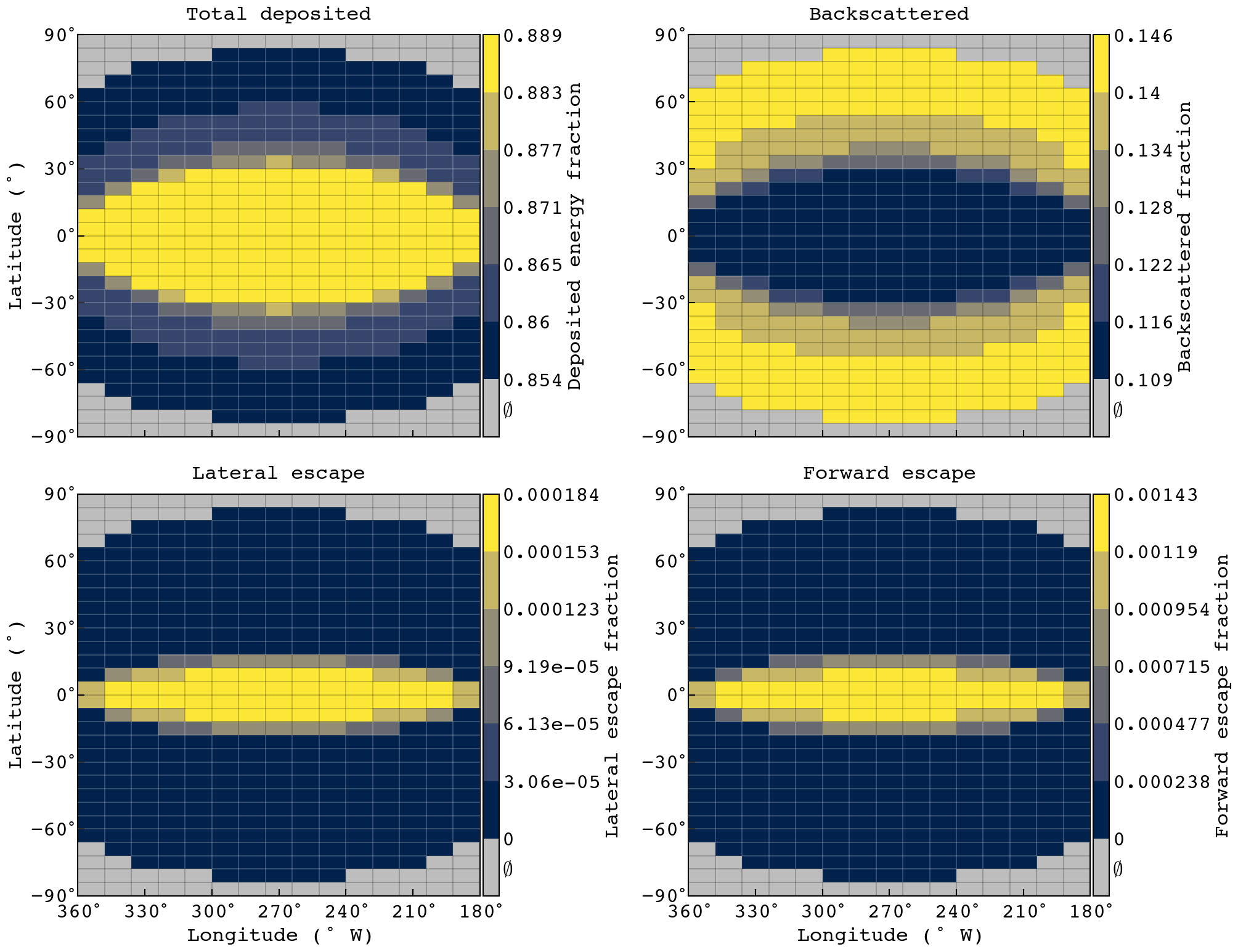}}
  \caption{Energy deposition diagnostics for electron bombardment simulations of Europa's trailing hemisphere.}
     \label{fig:europa_4p_trailing_edep}
\end{figure*}

\end{document}